
\magnification 1200
\hfuzz=3pt
\font\bb=msym10
\font\bbs=msym7
\def\e{E^{m,s}_\kappa }
\def\ez{E^{m,s}_0 }
\def\f{F^{m,s}_\kappa }

\def\s{\Sigma^{m,s}_\kappa}
\def\sz{\Sigma^{m,s}_0}
\def\fo{\varphi_0}
\def\fw{\varphi_\dbv}
\def\fww{\varphi_{\wdbv}}
\def\of{{\cal O}_{\varphi_0}}
\def\orb{{\cal O}^{m,s}_\kappa}

\def\opoin{{\cal O}_{P}(m, s)}

\def\lk{\lim_{\kappa\rightarrow 0}}

\def\hmsk{{\cal H}^{m,s}_\kappa}
\def\hmsz{{\cal H}^{m,s}_0}
\def\u{U^{m,s}_\kappa}
\def\h{{\cal H}}

\def\k{\kappa}

\def\dbv{{\hbox{\sevenrm w}}}
\def\wdbv{\tilde{\hbox{\sevenrm w}}}
\def\ddbv{{\hbox{\tenrm w}}}
\def\ddbvo{{\hbox{\tenrm w}}_{_{(0)}}}
\def\ww{\widetilde{\hbox{\tenrm w}}}

\def\ppfl{{\cal P}_+^\uparrow(3,1)}

\def\so{SO_0(3,2)}

\def\soo{so(3,2)}
\def\soc{so^{\cp}(3,2)}
\def\ssoo{so^*(3,2)}
\def\bzo{\bar z_{_{(0)}}}
\def\bxo{\bar \xi_{_{(0)}}}
\def\zo{z_{_{(0)}}}
\def\xo{\xi_{_{(0)}}}
\def\wo{\ddbv_{_{(0)}}}
\def\yo{y_{_{(0)}}}
\def\qo{q_{_{(0)}}}
\def\uo{u_{_{(0)}}}
\def\vo{v_{_{(0)}}}

\def\cz{\{5, 0, 1, 2, 3\}}
\def\yq{(y, q, u, v, t)}
\def\ky{\k Y}
\def\pk{\phi_\k}

\def\pu{\pi_1}
\def\pd{\pi_2}

\def\th{\theta^{\alpha\beta}}
\def\Y{Y_{\alpha\beta}}
\def\X{X_{\alpha\beta}}

\def\L{L_{\alpha\beta}}
\def\wL{\widetilde\L}
\def\hL{\hat L_{\alpha\beta}}

\def\T{\hbox{\bb T}}
\def\C{\hbox{\bb C}}
\def\R{\hbox{\bb R}}

\def\cp{\hbox{\bbs C}}
\def\rcinq{\R^5}

\def\rcinqeta{(\rcinq, \eta)}
\def\ome{\omega_{_E}}
\def\omf{\omega_{_F}}
\def\os{\omega_{_\Sigma}}

\def\ovk{\overrightarrow{\kern-2pt K\kern2pt}}
\def\ovp{\overrightarrow{\kern-2pt P\kern2pt}}
\def\ovs{\overrightarrow{\kern-2pt S\kern2pt}}
\def\ovj{\overrightarrow{\kern-2pt J\kern2pt}}
\def\ovx{\overrightarrow{\kern-2pt X\kern2pt}}
\def\te{\theta_{_E}}
\def\ke{K_{_E}}
\def\demi{{1\over 2}}

\def\msk{$(m, s, \k)$ }
\def\a{\`a }

\def\ab{\alpha, \beta \in \{5, 0, 1, 2, 3\}}
\def\mn{\mu, \nu \in \{0, 1, 2, 3\}}
\def\iu{i\in\{1, 2, 3\}}
\def\mk{{m\over\k}}
\def\kz{$\k\rightarrow0$}
\def\kzsd{\k\rightarrow0}

\def\picture #1 by #2 (#3){
        \vbox to #2{
             \hrule width #1 height 0pt depth 0pt
              \vfill
                                                        \special{picture #3}}}

\def\scaledpicture #1 by #2 (#3 scaled #4){{
        \dimen0=#1  \dimen1=#2
        \divide\dimen0 by 1000  \multiply\dimen0 by #4
        \divide\dimen1 by 1000  \multiply\dimen1 by #4
        \picture  \dimen0 by \dimen1 (#3 scaled #4)}}

\def\ref#1{\noindent\llap{[{\bf #1}]\quad}}
\def\refe#1{\item{\hbox to\parindent{\enskip[{\bf #1}]\hfill}}}

\def\page#1{\leaders\hbox to 5mm{\hfil.\hfil}\hfill
\rlap{\hbox to 5mm{\hfill#1}}\par}
\font\subsect=cmbx10 scaled \magstep1
\font\sect=cmbx10 scaled\magstep2
\font\titre=cmbx10 scaled \magstep3
\newif\ifpagetitre                          \pagetitrefalse
\newtoks\hautpagetitre                  \hautpagetitre={\hfil}
\newtoks\chapitrecourant               \chapitrecourant={\hfil}
\newtoks\titrecourant                     \titrecourant={\hfil}
\newtoks\hautpagegauche                \hautpagegauche={\hfil}
\newtoks\hautpagedroite
\hautpagedroite={\hfil\the\titrecourant\hfil}
\headline={\ifpagetitre\the\hautpagetitre
                  \else\ifodd\pageno\the\hautpagedroite
                  \else\the\hautpagedroite\fi\fi}
\footline={\hfil\bf\folio\hfil}

\def\makeheadline{\vbox to 0pt{\vskip -30pt
       \line{\vbox to8.5pt{}\the\headline}\vss}\nointerlineskip}
\def\makefootline{\baselineskip=35pt\line{\the\footline}}
\vsize=7.2 in
\voffset=0 mm
\hsize=138 mm
\baselineskip=15pt
\parskip=3pt plus 1pt minus 1pt


\line{October 23, 1992\hfill  CRM-1838}
\line{ \hfill }
\vglue .5cm
\centerline{\titre Phase Space Quantum Mechanics}
\smallskip
\centerline{\titre on the Anti-De Sitter Spacetime}
\smallskip
\centerline{\titre and its Poincar\'e
Contraction\footnote{$\,^\&$}{\it This work is largely based
on the PhD thesis of one of the authors (A.M.E.), presented}\footnote{}{\it
at Universit\'e Paris 7, december 1991.}}  \vglue 0.7cm

\centerline{\bf Amine M. El Gradechi\footnote{$\,^\ddagger\ $}{\it
e-mail: elgradec@ERE.UMontreal.CA}}  \vglue 0.1cm
{\baselineskip=12pt
\centerline{\it Centre de Recherches Math\'ematiques, Universit\'e de
Montr\'eal}   \centerline{\it C.P. 6128-A, Montr\'eal (Qu\'ebec) H3C 3J7,
Canada} \smallskip \centerline{\it and}
\smallskip
\centerline{\it Department of Mathematics, Concordia University}
\centerline{\it Montr\'eal (Qu\'ebec) H4B 1R6, Canada}\par}
\vglue 0.3cm
\centerline{\bf Stephan De Bi\`evre\footnote{$\,^\dagger\ $}{\it
e-mail: debievre@mathp7.jussieu.fr}}
\vglue 0.1cm
{\baselineskip=12pt
\centerline{\it UFR de Math\'ematiques and Laboratoire de Physique
Th\'eorique et Math\'ematique \footnote{\rm *\ }{\it C.N.R.S., ER
0004.}$\!\!$,} \centerline{\it Universit\'e Paris 7, T.C. 3\`eme \'etage, 2
place Jussieu F-75251 Paris Cedex 05, France}\par}
\vglue 0.5cm
\centerline{\bf ABSTRACT}
\vglue 0.2cm
{\baselineskip=13pt
In this work we propose an alternative description of
the quantum mechanics of a massive and spinning free particle
in anti-de~Sitter spacetime, using a phase space rather than a
spacetime representation.
The regularizing character of the curvature appears clearly in connection
with a notion of localization in phase space which is shown to
disappear in the zero curvature limit.  We show in particular how the
anti-de~Sitter optimally localized (coherent) states contract to plane
waves as the curvature goes to zero.  In the first part we give a detailed
description of the classical theory {\it \a la Souriau\/}.   This serves as
a basis for the quantum theory which is constructed in the second part using
methods of geometric quantization.  The invariant positive K\"ahler
polarization that selects the anti-de~Sitter quantum elementary system is
shown to have as zero curvature limit the Poincar\'e polarization which is
no longer K\"ahler.  This phenomenon is then related to the disappearance of
the notion of localization in the zero curvature limit.\par}

\vfill\eject

\noindent {\sect 1. Introduction}
\medskip
\noindent It is a well known fact that the Poincar\'e group,
$\ppfl$, the kinematical group of Minkowski spacetime, can be obtained by
means of a contraction from the anti-de~Sitter (AdS) group, $\so$, the
kinematical group of anti-de~Sitter spacetime [BLL] [LN].  The contraction
parameter is the constant positive curvature $\k$ of the anti-de~Sitter
spacetime.  This contraction procedure is thus nothing but a zero curvature
limit.  Accordingly, one would like to approximate
$\ppfl$-invariant theories by $\so$-invariant ones, hoping that such
approximations give rise to regularized relativistic theories.  Indeed, the
nonzero curvature equips the AdS theories with a lengthlike parameter,
which is the source of the sought for regularizations.

Up to now, this very stimulating idea has not been fully
exploited, though it has received a large amount of attention
for its potential implications in the context of quantum field theories
[BFFS].  We will concretely implement this idea in connection with the
problem of localization.  It is well known that no satisfactory notion of
space or spacetime localization in Poincar\'e-invariant quantum mechanics
exists [H1] [H2] [NW] [W] (see however [DB1]).  In the usual formulation of
relativistic quantum field theories, only momentum probability densities
are associated to the one-particle states.  In quantum theories on
anti-de~Sitter spacetime, no clear notion of localization has so far been
developed be it on spacetime, or in momentum space (for an attempt in
this direction see [Fr2]).  This makes the interpretation of one-particle
states very difficult.  We show in this work that the AdS quantum theory of
massive particles admits a very natural notion of {\it phase space\/}
localization.  In addition, we identify certain states of the theory as
optimally localized and show that they are -in a sense- the analogs of
plane waves  on flat spacetime.  We show in which sense the appearance
of the notion of phase space localization is a manifestation of the
regularizing character of the curvature.

More precisely, for a free massive and spinning particle in the
$4$-dimensional AdS spacetime, the phase space is a K\"ahler $\so$
homogeneous space, whose (geometric) quantization gives rise to a discrete
series representation of $\so$.  The latter is known to be a square
integrable representation, so the modulus of the
wave functions of the quantum states in this realization can be actually
interpreted as a probability distribution on phase space.  Moreover its
Hilbert space contains a particular family of quantum states: the Perelomov
[Pe] generalized coherent states.  They are the above mentioned {\it
optimally localized\/} states in phase space.  Here we exhibit the explicit
form of these coherent states and we show how their physical
interpretation arises.  We also stress the disappearance of this notion of
localization in the flat space limit, confirming the effectiveness of the
regularizing character of $\k$.

As a byproduct, the methods used here allow one to shed some light on the
problem of contracting discrete series representations.  The
case of the principal series representations was extensively studied using
different approaches [MN] [PW].  According to Mackey [Ma] the analogy
between these kind of representations, for a given (semi)simple noncompact
group $G$, and those of the semi-direct product group obtained through a
In\"on\"u-Wigner contraction [IW] of $G$, greatly simplifies the contraction.
The discrete series representations are far from possessing such an
analogy, and so their contraction is more difficult to analyze.  The
geometric quantization methods we use here give an idea of the kind of
difficulties one faces when dealing with the contraction of the discrete
series representations.  The observations made here and in [DBE] allowed
Cishahayo and De~Bi\`evre [CDB] to treat the $SU(1,1)\rightarrow {\cal
P}_+^\uparrow(1,1)$ case in a rigorous mathematical way.

We shall proceed as follows.  In section 2 we introduce the AdS spacetime
and we discuss some of its properties.  In section 3 we give a careful
and pedagogical geometric description ({\it \`a la} Souriau [So]) of the
classical theory of a mass $m\not=0$ and spin $s$ test particle in AdS
spacetime.  This fixes the physical interpretation of the different
quantities that will be used throughout this paper.  In the zero curvature
limit we recover the original Souriau geometric description of a mass
$m\not=0$ and spin $s$ test particle in Minkowski spacetime [So].  Using
methods of geometric quantization [Wo], the quantum theory is explicitly
constructed in section 4.  Section 5 deals with the notion of localization in
phase space.  In fact the latter is defined exploiting properties of the
generalized coherent states of $\so$.  The explicit form of those states is
derived there.  We also show that this notion of optimal localization
disappears in the zero curvature limit, relating this fact to the loss of the
K\"ahler character of the polarization in this same limit.  In section 6 we
explore the behaviour of the discrete series representation, explicitly
obtained in section 4, when the curvature tends to zero.  This provide us
with some information concerning the contraction of this type of
representations in the large.  Finally section 7 concludes our contribution.

The results presented here are more extensively discussed in an unpublished
thesis [E1].  The special case of a massive and spinless free particle
constitutes a straightforward generalization of the $1+1$-dimensional
case treated in [DBE].  It will not be considered here.

\bigskip
\noindent {\sect 2. The AdS
spacetime}
\medskip \noindent The AdS spacetime [DS] of (constant) curvature $\k>0$
can be viewed as the one sheeted hyperbolo\"\i d in $\rcinqeta$,
$\eta=diag(\buildrel 5\over{-}, \buildrel 0\over{-}, \buildrel 1\over{+},
\buildrel 2\over{+}, \buildrel 3\over{+})$, $$y\cdot y \equiv
\eta_{\alpha\beta} y^\alpha y^\beta = -(y^5)^2-(y^0)^2 +(y^1)^2 + (y^2)^2 +
(y^3)^2 = -\k^{-2},  \eqno (2.1)$$ where $\alpha,\beta \in \lbrace
5,0,1,2,3\rbrace$ and $(y^5, y^0, y^1, y^2, y^3) \in \rcinq$.  In what follows
we shall also denote this spacetime by $M_\k$.  Figure 2.1 below displays
the two dimensional version of $M_\k$.

\vglue 6cm
\nobreak\centerline{\bf Figure 2.1}
\centerline{\it Anti-de~Sitter spacetime\/}

Alternatively, $M_k$ can be realized through global
coordinates $(x^0, \vec x)$. The latter are related to those in (2.1) by the
following relations,
$$y^5=Y\cos \k x^0,$$ $$y^0=Y\sin \k x^0,\eqno(2.2a)$$ $$\vec y = \vec
x,$$   where $\quad-\pi\leq\k x^0\leq\pi$, $\enskip \vec x \in
\R^3\enskip$  and
$$Y=\sqrt{\k^{-2}+(\vec x)^2}.\eqno(2.2b)$$
In this coordinate system the metric on $M_\k$ takes the form,
$$\openup 2mm\eqalignno{ds_\k{}^2
&=g_{\mu\nu}\,dx^\mu\, dx^\nu\cr
&=-(\k Y)^2\,(dx^0)^2+(\k Y)^{-2}\,dr^2+r^2\left(d\theta^2+\sin^2\theta\,
d\phi^2\right).&(2.3)}$$
Here $(r, \theta, \phi)$ are the usual spherical coordinates in
$\R^3$.  Obviously $x^0$ is the time coordinate, $\k x^0$ being
the rotation angle in the $(y^5, y^0)$ plan of (2.1) and $\vec x$ or $\vec y$
are the usual $3$-space coordinates.

Let us mention that the known problems inherent to the intrinsic geometry
of $M_\k$, namely the compactness of time and the absence of global
hyperboliticity, will not be addressed here.  In fact, the first
problem can be avoided by considering the AdS spacetime to be the
universal covering of $M_\k$ [Fr2] [HE].  A careful study of the second can be
found in [AIS], where suggestions for its resolution have also been
formulated (see also [Fr2] and [CB]).

The isometry group of the AdS spacetime appears clearly from (2.1) to be
the non-compact $O(3,2)$ group.  The connected component to the identity
of the latter, denoted $\so$, is called the AdS group [HE].  Together with
the $SO_0(4,1)$-de~Sitter spacetime, $M_\k$ is the only non-trivial
maximally symmetric solution of Einstein equations (with non zero
cosmological constant).

Since in this work we are interested in carrying out a
zero curvature limit, it is worth noting that this procedure will actually
produce physically relevant quantities provided it is performed using a
meaningful parametrization.  For example, in the case of $M_\k$, one can
see that contrary to the $y$-coordinates in (2.1), the $x$-coordinates in
(2.2) possess a straigthforward interpretation for any value of $\k$ and
even  when \kz. In fact, a simple \kz\ limit in (2.3) shows that
$ds_k{}^2$ becomes the Minkowski flat metric.  Hence one can still
interpret $(x^0, \vec x)$ as the spacetime coordinates.  In other words the
$y$ and the $x$-coordinates are complementary for the purpose of the
present work.  The former put in perspective the symmetry of the system,
their transformation under the action of $\so$ being obvious.  The latter
provide the bridge towards a meaningful zero curvature limit, yielding the
expected physical quantities.

At the algebraic level, both the Poincar\'e and AdS Lie algebras have an
underlying ten dimensional vector space. The Poincar\'e Lie algebra can be
obtained as the limit of a one parameter-dependent sequence of isomorphic
AdS Lie algebras.  According to In\"on\"u and Wigner [IW] this singular
process is called a contraction.  The parameter used in this limiting process
can easily be seen to be the curvatutre $\k$ of the AdS spacetime. So this
contraction is nothing but the zero curvature limit mentionned above [BLL].

In order to fix the notations, let us explicitly perform the
AdS $\rightarrow$ Poincar\'e contraction.  Let $V\equiv \R^{10}$ be the
vector space underlying the two Lie algebra structures $p(3,1)$ and $\soo$
($\dim p(3,1)=\dim \soo=10$). Let  $\{e_{\alpha\beta}, \alpha, \beta  \in
\{5, 0, 1, 2, 3\}\}$ be a basis of $V$ such that $\soo$ is realized in the
following way,
$$\lbrack
e_{\alpha\beta},
e_{\gamma\rho}\rbrack=\eta_{\alpha\gamma}e_{\beta\rho}+
\eta_{\beta\rho}e_{\alpha\gamma}-\eta_{\alpha\rho}e_{\beta\gamma}-
\eta_{\beta\gamma}e_{\alpha\rho}.\eqno(2.4)$$
Notice that
$e_{\mu\nu}, \mn$ realize the Lorentz subalgebra $so(3,1)\subset\soo$.
Let $\pk\in GL(V)$ be the
{\it contraction map\/} [Do] defined by, $$\pk: V\ni
e_{\alpha\beta}\,\longmapsto\,\pk(e_{\alpha\beta})\equiv
e_{\alpha\beta}^\k,\eqno(2.5a)$$
where
$$e_{5\mu}^\k=\k
e_{5\mu},\eqno(2.5b)$$and$$e_{\mu\nu}^\k=e_{\mu\nu},\eqno(2.5c)$$
for $\mn$.  As long as $\pk$ is non-singular, i.e. $\k\not=0$, one can
define a new Lie algebra strucutre $\lbrack\ ,\ \rbrack_\k$ in $V$, which is
isomorphic to the original one.  Specifically,   $$\lbrack e_{\alpha\beta},
e_{\gamma\rho}\rbrack_\k=\pk^{-1}\lbrack \pk(e_{\alpha\beta}),
\pk(e_{\gamma\rho})\rbrack.\eqno(2.6)$$ However when $\k$ reaches $0$
one obtains a new Lie algebra structure which is no longer isomorphic to
$\soo$.  This is the Poincar\'e Lie algebra $p(3,1)$. Indeed by taking the
\kz\ limit in (2.6) we get, $$\lbrack e_{\alpha\beta},
e_{\gamma\rho}\rbrack_0=\lk\pk^{-1}\lbrack \pk(e_{\alpha\beta}),
\pk(e_{\gamma\rho})\rbrack,\eqno(2.7a)$$  which results in the following
explicit commutation relations of $p(3,1)$: $$\eqalignno{\lbrack
e_{\mu\nu},
e_{\delta\lambda}\rbrack_0&=\eta_{\mu\delta}e_{\nu\lambda}+
\eta_{\nu\lambda}e_{\mu\delta}-\eta_{\mu\lambda}e_{\nu\delta}-
\eta_{\nu\delta}e_{\mu\lambda},\cr \lbrack e_{\mu\nu},
e_{5\lambda}\rbrack_0&=\eta_{\mu\lambda}e_{5\nu}-
\eta_{\nu\lambda}e_{5\mu},&(2.7b)\cr \lbrack e_{5\mu},
e_{5\nu}\rbrack_0&=0.\cr}$$ Here $\mu, \nu, \delta, \lambda \in \{0, 1, 2,
3\}$.  The subalgebra $so(3,1)$ is clearly preserved in this
contraction, and $\so$ is then said to be contracted to $\ppfl$ along the
Lorentz subgroup $SO_0(3,1)$.  For more details concerning the procedure of
contraction we refer to [IW] [Sa] [Do] [LN] [Gi] (for more recent
contributions see [CPSW] and references quoted therein.)

\vfill\break
\noindent {\sect 3. The classical theory}
\medskip \noindent
As mentioned in the introduction, the theory we are about to develop here
describes the free evolution of a mass $m$ and spin $s$ particle on an AdS
spacetime of given curvature $\k>0$. Both the classical and the quantum
aspects of this theory will be investigated.  Here we concentrate on the
classical dynamics in order to fix both the notations and the physical
interpretation of the quantities that will appear throughout this work. The
quantum theory and the associated notion of optimal localization will be
discussed in the next sections.

As in [DBE] we shall use here the Souriau scheme [So] (see also [DB2]),
which provides the classical dynamics, its phase space and its symmetries
in a unified and an efficient way.  This scheme can be summarized in the
following diagram:

\vglue 4cm
\centerline{\bf Figure 3.1}
\centerline{\it Souriau's scheme\/}

The phase space $\Sigma$ of the model is obtained by a symplectic
reduction of a presymplectic manifold $E$.  Hence, the degenerate closed
two-form $\ome$ equipping $E$ is the pull-back of the
non-degenerate closed two-form $\os$ equipping $\Sigma$,
i.e. $\ome=\pi^*\os$.  In other words the projection $\pi$ kills the
kernel of $\ome$.  The presymplectic manifold $E$ is chosen in such a way
that the projection $\rho$ on $M$ of the leaves of the foliation generated
by the distribution $\ker\ome$ gives rise to the dynamics of the theory.
More precisely, in the case of a free massive particle on $M$ this should
produce the time-like geodesics of $M$. Unfortunately there exists no
general theory prescribing the choice of the presymplectic manifold $E$.
However the symmetries of the model, if any, provide a precious guide
for such a determination.  Let us moreover anticipate by indicating that
even if the above scheme concerns just the classical theory, it
highly simplifies the quantization procedure.  This will be shown in section
4.

In what follows we shall sometimes use Souriau's terminology. The
presymplectic manifold and the phase space will then be called the
{\it evolution space\/} and the {\it space of motions\/}, respectively.
\vfill\break
\noindent{\subsect 3.1. Souriau's scheme}
\medskip
\noindent For the present case the evolution space, denoted $\e$, will be
taken as a principal homogeneous space [LM] of $\so$, i.e. $\e\cong\so$. It
can be concretely and conveniently realized as a subspace of the cartesian
product of five copies of $\rcinqeta$.  Actually, let $\yq$ be five
five-vectors of $\rcinqeta$, then $\e$ is defined as the set of points
$\ddbv=\yq\in\R^{25}$ satisfying the following $\so$-invariant
constraints:

$$y\cdot
y =-\k^{-2}, \eqno (3.1a)$$  $$q\cdot q = -m^2, \eqno (3.1b)$$ $$u\cdot u =
1, \eqno (3.1c)$$ $$v\cdot v = 1, \eqno (3.1d)$$
$$t\cdot t = m^2s^2, \eqno (3.1e)$$
$$y\cdot q = y\cdot u=y\cdot v=y\cdot t=q\cdot u=q\cdot v=q\cdot t=u\cdot
v=u\cdot t=v\cdot t=0, \eqno (3.1f)$$
$$\epsilon _{\alpha \beta \gamma \rho \sigma}\, y^\alpha q^\beta u^\gamma
v^\rho t^\sigma = {m^2s \over \kappa}, \eqno (3.1g).$$
and
$$y^5q^0-y^0q^5>0.\eqno(3.1h)$$
In these equations $m$, $s$ and $\k$ are the three original physical
ingredients and $\epsilon _{\alpha \beta \gamma \rho  \sigma}$ is the
completely skew-symmetric tensor associated to $\rcinqeta$, such that
$\epsilon_{50123}=\epsilon^{50123}=1$. The indices of the vectors $\yq$
are raised and lowered by $\eta_{\alpha\beta}$ (2.1).  Note also that $q$,
$u$, $v$ and $t$ will be considered either as points of the tangent or the
cotangent space to $M_\k$ (2.1).

The physical interpretation of the above constraints is now displayed in
some detail:

\noindent\itemitem{$\bullet\ $(3.1a)\enskip} {defines the AdS spacetime
points $y\in M_\k$ (2.1);}

\noindent\itemitem{$\bullet\ $(3.1b)\enskip}{from $y\cdot q=0$ in (3.1f)
$q$ appears as the conjugate linear momentum of the position $y$,
through (3.1b) it is constrained to the {\it AdS mass shell\/} associated
to $m$;}

\noindent\itemitem{$\bullet\ $(3.1e)\enskip}{$t$ is what we call the
{\it AdS-Pauli-Lubanski\/} five-vector; (3.1e) is the AdS analog of the
Pauli-Lubanski constraint appearing in the case of the Poincar\'e-invariant
theory [So], and this is the way the spin enters in our approach; note that
$t$ is spacelike, belongs to $T_yM_\k$ and is perpendicular to the direction
of motion (3.1f);}

\noindent\itemitem{$\bullet\,$(3.1c,d)}{these two constraints allow a
covariant treatment of spin and they will play an important role at the
quantum level; we shall see that $t$ and $(u,v)$ are equivalent;}

\noindent\itemitem{$\bullet\,$(3.1g,h)}{these constraints specify a choice of
orientation and select one of the four connected components of the manifold
defined just by the previous constraints (3.1a-f); in particular (3.1h)
selects one of the two mass shells satisfying (3.1b), in the zero curvature
limit this corresponds to the positive energy condition.

Since one of our goals is to perform the zero curvature limit, we will
introduce, later on in section 3.2, a new coordinate system that will make
such a procedure both possible and meaningful. This is related to the
remarks we made in section 2 concerning the complementarity of the $y$ and the
$x$-coordinates. The preceding constraints will then be rewritten in that
new coordinate system and their zero curvature limits will be evaluated, the
results we will obtain will confirm the physical interpretation we gave
above.  At this point it is worth noting that $\e$ can be viewed as the
{Lorentz bundle\/} over the AdS spacetime. This makes clear the connection
with K\"unzle's 1972 work [Ku].  There the evolution space for a free
massive and spinning particle on a general spacetime is taken as the
Lorentz bundle over that spacetime.

Simple arguments of linear algebra allow one to make the identification
$\e\cong\so$. Concretely, to each $\ddbv\in\e$ we can associate in a $1$-$1$
manner an $\so$ element. We must first fix the point $\ddbvo$ that is
associated to the identity element of $\so$. We choose
$\wo\equiv(y_{_{(0)}}, q_{_{(0)}}, u_{_{(0)}}, v_{_{(0)}}, t_{_{(0)}})$ as
follows:
$$y^\alpha_{_{(0)}} = \kappa^{-1}\, \eta^{\alpha}{}_5,\quad q^\alpha_{_{(0)}}
= m\,
\eta^{\alpha}{}_0,\quad u^\alpha_{_{(0)}} = \eta^{\alpha}{}_1,\quad
v^\alpha_{_{(0)}} =
\eta^{\alpha}{}_2,\quad t^\alpha _{_{(0)}} = ms\, \eta^{\alpha}{}_3. \eqno
(3.2)$$
{}From here one can identify the $\so$ element $\Lambda(\ddbv)$ associated
to a general point $\ddbv$ of $\e$.  In fact
$\Lambda(\ddbv)$ is the group element relating $\ddbvo$ to $\ddbv$ when
$\so$ acts on $\e$ on the left.  It is given by,
$$\big[
\Lambda(\ddbv)\big]^{\mu}{}_\nu = -\kappa^2 y^\mu y_{(0)\nu}-{1\over
{m^2}} q^\mu q_{(0)\nu} + u^\mu u_{(0)\nu} + v^\mu v_{(0)\nu} + {1\over
{m^2s^2}} t^\mu t_{(0)\nu}. \eqno (3.3)$$

The Lie algebra $\soo$ in (2.4) can now be realized in terms of (left)
invariant vector fields on ($\so\equiv$)$\e$.  Their expressions are obtained
from the formula below,
$$\openup 3mm\displaylines{Y_{\alpha\beta}(\ddbv)= {d\over
d\tau}\left[\lbrack\Lambda(\ddbv)\rbrack^\mu{}_\nu(\exp\,\tau
e_{\alpha\beta})^\nu{}_\rho\,
y_{_{(0)}}^\rho\right]_{\tau=0}\!{\partial\over\partial y^\mu}+ {d\over
d\tau}\left[\lbrack\Lambda(\ddbv)\rbrack^\mu{}_\nu(\exp\,\tau
e_{\alpha\beta})^\nu{}_\rho\,
q_{_{(0)}}^\rho\right]_{\tau=0}\!{\partial\over\partial q^\mu} \hfill\cr
\hfill+{d\over
d\tau}\left[\lbrack\Lambda(\ddbv)\rbrack^\mu{}_\nu(\exp\,\tau
e_{\alpha\beta})^\nu{}_\rho\,
u_{_{(0)}}^\rho\right]_{\tau=0}\!{\partial\over\partial u^\mu} +{d\over
d\tau}\left[\lbrack\Lambda(\ddbv)\rbrack^\mu{}_\nu(\exp\,\tau
e_{\alpha\beta})^\nu{}_\rho\,
v_{_{(0)}}^\rho\right]_{\tau=0}\!{\partial\over\partial v^\mu} \!\!\!\!\!\cr
\hfill+{d\over
d\tau}\left[\lbrack\Lambda(\ddbv)\rbrack^\mu{}_\nu(\exp\,\tau
e_{\alpha\beta})^\nu{}_\rho\,
t_{_{(0)}}^\rho\right]_{\tau=0}\!{\partial\over\partial t^\mu}.
\qquad\qquad\qquad\qquad\qquad(3.4)\cr}$$
Using (3.3), (3.2) and the equation $\big\lbrack{d\over d\tau}(\exp\,\tau
e_{\alpha\beta})\big\vert_{\tau=0}\big\rbrack^\nu{}_\rho=\eta_{\alpha\rho}
\eta^\nu{}_\beta-
\eta_{\beta\rho}\eta^\nu{}_\alpha$ we get the $\Y$'s explicitly as
follows,
$$Y_{50}=m\kappa\, y\cdotp{\partial\over\partial q}-{1\over
m\kappa}\,q\cdotp {\partial\over\partial y},\qquad Y_{12}= v\cdotp
{\partial\over\partial u}-u\cdotp{\partial\over\partial v},\eqno(3.5a)$$
$$Y_{23}={1\over ms}\, t\cdotp{\partial\over\partial v}-ms\,v\cdotp
{\partial\over\partial t},\quad Y_{31}= ms\,u\cdotp{\partial\over\partial
t}-{1\over ms}\, t\cdotp{\partial\over\partial u},\eqno(3.5b)$$
$$Y_{15}={1\over \kappa}\, u\cdotp{\partial\over\partial
y}+\kappa\,y\cdotp {\partial\over\partial u},\ \  Y_{25}={1\over
\kappa}\,v\cdotp{\partial\over\partial y}+\kappa\,
y\cdotp{\partial\over\partial v}, \ \  Y_{35}={1\over ms\kappa}\,
t\cdotp{\partial\over\partial y}+ms\kappa\,y\cdotp {\partial\over\partial
t},\eqno(3.5c)$$ $$Y_{01}=-m\, u\cdotp{\partial\over\partial q}-{1\over
m}\,q\cdotp {\partial\over\partial u},\ \  Y_{02}=
-m\,v\cdotp{\partial\over\partial q}-{1\over m}\,
q\cdotp{\partial\over\partial v}, \ \  Y_{03}=-{1\over s}\,
t\cdotp{\partial\over\partial q}-s\,q\cdotp {\partial\over\partial
t}.\eqno(3.5d)$$
\medskip
\noindent At each point $\ddbv\in\e$, these vector fields are
linearly independent, they form at this point a basis of
$T_{_\dbv}\e$.

The identification $\e\cong\so$ just realized plays a crucial
role in the present construction.  In particular it allows us to choose an
invariant presymplectic form $\ome$ in an easy way.  In order to show this,
let us first introduce the dual basis to $\{\Y,
\ab\}$ denoted  $\{\th, \ab\}$.  The $\th$'s are (left) invariant one-forms
on $\e$ and can be viewed as the basis elements of $\ssoo$.  Then
$\ome$ can be chosen as the exterior derivative of an invariant
one-form $\te$, which is some linear combination of the $\th$'s with constant
coefficients.  The two-form $\ome\equiv-d\te$ will then
be closed and invariant as wanted.  But still one needs to make a choice
among all possible linear combinations.  The only requirement really
constraining this choice is that $\ker\ome$ should produce the right
dynamics on $M_\k$.  We will choose here $\te$ on the basis of some
physical arguments, and we will confirm the validity of this
choice subsequently by the evaluation of the dynamics.

We will use here dimensionality and kinematic arguments.  First of all,
$\te$ must have the dimension of an action.  Its expression must contain
the three original physical ingredients, namely the mass $m$, the spin $s$
and the curvature $\k$.  Since the $\th$'s are dimensionless, the
dimensionality  of $\te$  can only arise from $m$, $s$ and $\k$.  One can
easily check that only $\mk$ and $s$ have the dimension of an action (in
$\hbar=c=1$ units).  On the other hand, it is well known that the mass and
the spin are kinematically related to the spacetime translations and space
rotations, respectively.  Since we are dealing here with the $\so$
kinematics, the spacetime translations are pseudo-rotations, with
dimensionless parameters.  The latter acquire actual lenghthlike
dimension when multiplied by $\k^{-1}$, the unit of length [BEGG].  All these
arguments put together, suggest the folowing choice for $\te$,
$$\te\equiv{m\over\kappa}\,\theta^{50}+s\,\theta^{12}.\eqno(3.6)$$

At this point we must mention that even if this choice gives the right
dynamics, as we will soon show, it still carries some arbitrariness.  The
latter is inherent to classical theory and can be compared to the one
appearing in the Lagrangian formalism when one tries to fix a Lagrangian.
What is important in that formalism is that the equations of motions
describing the dynamics of the system under study are obtained correctly
from the chosen Lagrangian via the variational principle.

In order to write (3.6) in a concrete form, let us first find the $\th$'s in
terms of the coordinates on $\e$.  This is done using (3.5) and the duality
relation,
$$Y_{\gamma\rho}\,\rfloor\theta^{\alpha\beta}\equiv\theta^{\alpha\beta}
(Y_{\gamma\rho})=
\eta^\alpha{}_\gamma\eta^\beta{}_\rho-\eta^\alpha{}_\rho\eta^\beta{}_\gamma,
\qquad
\forall\, \alpha, \beta, \gamma, \rho \in \{5, 0, 1, 2, 3\}. \eqno(3.7)$$
We display here only the results that will be needed subsequently,
$$\theta^{50}={\kappa\over m}\,q\cdot
dy=-{\kappa\over m}\,y\cdot dq\qquad {\rm and}\qquad\theta^{12}= v\cdot
du=-u\cdot dv;\eqno(3.8)$$
note that (3.1f) is at the origin of the second equalities in (3.8).  Finally,
we
obtain for (3.6)$$\te=q\cdot dy+s\,v\cdot du.\eqno(3.9)$$
The expression of $\te$ in (3.6) has a group theoretic character.  Its
translation in geometric terms given in (3.9) is based, once again, on the
identification $\e\cong\so$, and it allows us to view it as the restriction to
the case of a constant curvature manifold of the one-form used in [Ku].

Let us now give the expression of the invariant presymplectic form $\ome$,
$$\ome\equiv-d\theta_E=dy\wedge dq+s\,du\wedge dv;\eqno(3.10)$$
the convention used here is $dy\wedge dq\equiv dy^\alpha\wedge
dq_\alpha$, where Einstein's summation rule is assumed.  The canonical
character of (3.10) confirms the physical interpretation we assigned to
the coordinates $\yq\in\e$.  Another useful formula for $\ome$ can be
obtained starting with (3.6) and using the Maurer-Cartan
equations of $\ssoo$.  Hence,
$$\ome=\mk\,
(\theta^{01}\wedge\theta^{15}+\theta^{02}\wedge\theta^{25}+
\theta^{03}\wedge\theta^{35})-s\,(\theta^{15}\wedge\theta^{25}+\theta^{01}\wedge
\theta^{02}+
\theta^{31}\wedge\theta^{23}).\eqno(3.11)$$

In order to complete the description of Souriau's scheme, we come now to
the evaluation of the kernel of $\ome$.   The latter is defined as follows,
$$\ker
\ome=\bigcup_{\dbv\in\e}\ker_{_\dbv}\ome\eqno(3.12a)$$  where
$$\ker_{_\dbv}\ome= \{Y \in
T_{_\dbv}\e\bigm\vert\ome(\ddbv)\big(Y,Y'\big)=0\quad\forall\ Y'\in
T_{_\dbv}\e\}.\eqno(3.12b)$$
Evaluating
$\ker\ome$ is equivalent to solve the equation,
$$\ke\, \rfloor\ome=0,\eqno(3.13)$$
for $\ke$ a $\ker\ome$-valued vector field on $\e$, i.e. $\ke(\ddbv)\in
\ker_{_\dbv} \forall \ddbv \in \e$.  The symbol ``$\rfloor$" denotes the
interior product.  Since the $\Y$'s form a basis of $T_{_\dbv}\e$ at each
point $\ddbv\in\e$, $\ke$ can be expanded in the following way,
$$\ke=\demi
f^{\alpha\beta}\Y,\eqno(3.14)$$ with $f^{\alpha\beta} \in
C^\infty(\e)$.  Using (3.14) and (3.11) it is easy to solve
(3.13).  A very interesting situation appears.  In fact two cases must
be considered. They are summarized as follows:

\item{\bf(i)}{$\displaystyle\mk=s\Longrightarrow \dim\ker\ome=4$,
$\ker\ome$ is spanned by four linearly independent vector fields on
$\e$;}
\item{\bf(ii)}{$\displaystyle\mk\not=s\Longrightarrow \dim\ker\ome=2$,
$\ker\ome$ is spanned by two linearly independent vector fields on $\e$.}

We consider in this work only the second case, the first one deserves a
separate treatment and it will be discussed elsewhere [E2].  However let us
say a few words about this phenomenon.  For the case (i) the symplectic
reduction $\pi: \e\longrightarrow\e/\ker\ome$ will yield a
six-dimensional phase space.  In the case of a $\ppfl$-invariant free
theory, six-dimensional phase spaces describe either massive spinless
free particles or massless spinning ones [Ar] [So].  Since the Lorentz
subgroup of $\so$ is preserved when contracting, the spin part of any
physical $\so$-invariant theory will also be preserved in the zero
curvature limit.  So the second possibility above, namely mass $=0$ and
spin $\not=0$, is the only one that could arise from a zero curvature limit
starting from the six-dimensional $\so$-phase space corresponding to
the case $\mk=s$.  Thus, in a sense to be defined, the case (i)
corresponds to an $\so$-invariant free {\it massless\/} elementary
system.  For more details we refer to [E2].

The symplectic reduction in case (ii), which is our main concern here, gives
rise to an eight-dimensional phase space.  For a $\ppfl$-invariant free
theory, only a massive
spinning free particle on Minkowski spacetime is described by an
eight-dimensional phase space.  So, as we will show subsequently,  the
latter is the zero curvature limit of the former.  Or the other way around,
the former is the AdS-deformation of the latter.  This suggests calling
the $\mk\not=s$ case, the {\it massive\/} case.

When $\mk\not=s$, one can show that $\ker\ome$ is spanned by the two
vector fields $Y_{50}$ and $Y_{12}$ given in (3.5a).  Hence, $\ke$ is of the
form, $$\ke=f^{50}Y_{50}+f^{12}Y_{12},\eqno(3.15)$$
for $f^{50}$ and $f^{12}$ arbitrary elements of  $C^\infty(\e)$.  Recall now
that $Y_{50}$ and $Y_{12}$ together, as $\soo$ basis elements, generate the
subgroup $SO(2)\times SO(2)$.  As vector fields on $\e$ they have closed
integral curves.  The integral manifold or the
leave of $\ker\ome$ through each point
$\ddbv\in\e$ is then a torus $\T=S^1\times S^1$.  The symplectic
reduction allows then to identify the phase space of our physical system as
$\s\equiv\e/S^1\times S^1\cong\so/SO(2)\times SO(2)$.  Hence, $\s$ is a
homogeneous space of $\so$.  Moreover, as we will show in section 4, $\s$
is a K\"ahler homogeneous space of $\so$.  The symplectic reduction
stressed here can be explicitly carried out, namely $\s$ can be given
coordinates with clear physical interpretation.  Formulas with this respect
will be displayed later on.

In order to evaluate the dynamics of our system let us now make a short
detour and complete Souriau's scheme.  To this end we must first
integrate $\ker\ome$ and then project its leaves  on $M_\k$.  Just by
looking at the form of $Y_{50}$ and $Y_{12}$ in (3.5a) one can easily see
that only the integral curves of $Y_{50}$ project non-trivially on
$M_\k$, giving birth to worldlines.  The equations of the latter result
from the integration of the flows of $Y_{50}$.  Concretely, this gives the
following solution,
$$\left\{\normalbaselineskip=24pt\matrix{y^\alpha(\tau)=y^\alpha(0)\,\cos\tau-\d
isplaystyle
{1\over m\k}\, q^\alpha(0)\,\sin\tau\cr
q^\alpha(\tau)=m\k\,y^\alpha(0)\,\sin\tau+q^\alpha(0)\,\cos\tau\hfill\cr}
\right.
\qquad \alpha\in\cz,\eqno(3.16)$$
$y^\alpha(0)$ and $q^\alpha(0)$ are the initial conditions and
$\tau\in\R$.  It is easy to see that for each such initial condition,
$y^\alpha(\tau)$ traces out a timlelike geodesic on $M_\k$.
Hence, our choice for $\te$ in (3.6) generates the expected dynamics
of our free AdS system.  Finally, note that the position and the spin degrees
of freedom behave in an independent way.  This is a consequence of the
fact that $Y_{50}$ and $Y_{12}$ commute.

Let us now come back to the symplectic reduction sketched above.
Concretely, we define the phase space $\s$ in the following way,
$$\s=\Big\{\,(y,
q, t)\in\f\mid y^0=0\ \hbox{and}\ y^5>0\Big\}.\eqno(3.17)$$
Here $\f$ is a $9$-dimensional manifold obtained through the partial
reduction of $\e$:
$$\pu: \e\ni(y, q, u, v,
t)\,\longmapsto\,(y, q, t)\in\f.\eqno(3.18)$$
Clearly $\f$ is a submanifold of the cartesian
product of three copies of $\rcinqeta$, $\f\subset\R^{15}$.
This manifold is presymplectic, its presymplectic form $\omf$ is obtained
from the pull-back:
$$\ome=\pu{}^*\omf.\eqno(3.19)$$
More explicitly,
$$\omf=dy\wedge
dq-{\k\over2m^2}\,\epsilon_{\alpha\beta\gamma\rho\sigma}\,y^\alpha
q^\beta t^\gamma \Big\lbrack\k^2\, dy^\rho\wedge dy^\sigma+{1\over
m^2}\, dq^\rho\wedge dq^\sigma-{1\over m^2s^2}\, dt^\rho\wedge
dt^\sigma\Big\rbrack.\eqno(3.20)$$
Moreover $\f$ is a homogeneous space of $\so$, namely
$$\f\cong\so/SO(2).\eqno(3.21)$$
Hence, the reduction $\pu: \e\,\rightarrow\,\f$
partially kills $\ker\ome$.  It kills the spin part represented by the flows
of the vector field $Y_{12}$.  The symplectic reduction process is
completed by killing $\ker\omf$.  This is actually done through choosing a
section of the bundle $\pd:\f\,\rightarrow\,\s\equiv\f/\ker\omf$. The
same choice as in [DBE] is made here (see (3.17)).  It's a simple one from the
computational point of view.  Each leaf of $\ker\omf$, when projected on
$M_\k$, gives rise to a timelike geodesic of $M_\k$ (as previously
described for $\ker\ome$).  This leaf can be uniquely represented by the
inverse image of the point of the geodesic that intersects the half plane
$y^0=0, y^5>0$.  Note that the timelike geodesics of $M_\k$ are all closed
curves [Wi2].  Physically this choice of section corresponds to choosing
initial conditions at time zero ($y^0=0\Leftrightarrow x^0=0$ see
(2.2a)).  The symplectic form $\os$ is such that $\omf=\pd^*\os$ or
equivalently $\ome=\pi^*\os$, where $\pi=\pd\circ\pu$.  More precisely,
$$\os=\omf\vert_{y^0=0}.\eqno(3.22)$$

At this point a precision is in order.  One could argue that we should have
started the construction by considering $\f$, instead of $\e$, as the
evolution space.  One can easily see from (3.20) that this is not an easy task.
In fact $\omf$ can not be easily guessed, whereas constructing $\ome$
was relatively straightforward.

We end this subsection by summarizing Souriau's scheme in a diagram.

\vglue 5cm
\centerline{\bf Figure 3.2}
\centerline{\it Souriau's scheme for the AdS massive spinning free
particle\/}

\bigskip

\noindent{\subsect 3.2. Contraction adapted coordinates}
\medskip
\noindent In this subsection we introduce new coordinates on
$\e$.  They are needed in order to concretely carry out the zero
curvature limit.  Up to now we used coordinates based on the
$y$-coordinatization of $M_\k$, the ones we introduce here are
based on the $x$-coordinatization of $M_\k$.
Thus, in the same way as the position $x$ replaces the
position $y$ (see(2.2)),  the linear momentum $p$,
the spin $s$ and the $a$ and $b$
four-vectors will replace $q$, $t$, $u$ and $v$,
respectively.  Subsequently we shall call the old and the new
coordinates on $\e$, the $y$ and the $x$-coordinates,
respectively.  One obtains relations linking old and new
coordinates, as in (2.2), through equations of the type:
$$q\cdot dy\equiv p_\mu\cdot dx^\mu=g_{\mu\nu}(x)p^\mu
dx^\nu,\eqno(3.23)$$
and their analogs for the other pairs of coordinates.  In (3.23) $g$
is the metric on $M_\k$ given in (2.3).
We will here only display the solutions of (3.23) for the $(q, p)$
pair, the others arise through obvious modifications.  We obtain,
$$\eqalignno{p_0&=\k\,\bigl(y^5q_0-y^0q_5\bigr),&(3.24a)\cr
p_i&=q_i+\bigl(\vec y\cdot\vec q\bigr)\,{y^i\over Y^2},
\qquad\iu;&(3.24b)\cr}$$ here $\vec y\cdot\vec
q=y^iq_i=\Sigma_{i=1}^3y^iq_i$.  When inverting (3.24a-b)
one gets,
$$\eqalignno{q_5&=-\k\,(\vec x\cdot\vec
p)\, (\k y^5)-{p_0\over(\ky)^2}\,(\k y^0),&(3.25a)\cr
q_0&=-\k\,(\vec x\cdot\vec p)\, (\k y^0)+{p_0\over(\ky)^2}\,(\k
y^5),&(3.25b)\cr q_i&=p_i+\k^2\,(\vec x\cdot\vec p)\,
x^i,\quad\iu;&(3.25c)\cr}$$ here $\vec
x\cdot\vec p=x^ip_i$.  Note also the interesting relation $\vec
y\cdot\vec q=(\ky)^2\,\vec x\cdot\vec p$.

 From now on we will reexpress the most important equations
derived in the previous subsection in terms of the $x$-coordinates
and then we will investigate their zero curvature limit.  We start
with the constraints defining $\e$ in (3.1a-h).  We present the
results in the following form,
$$\normalbaselineskip=24pt\matrix{y\cdot
y=-\k^{-2}&\longrightarrow& x\in M_\k \qquad\enskip\,
\qquad& \buildrel \kzsd\over{\longrightarrow}& x\in M_0\qquad
&\hbox{Minkowski spacetime}\ \cr    q\cdot
q=-m^{2}&\longrightarrow& g_{\mu\nu}(x)p^\mu
p^\nu=-m^{2} & \buildrel \kzsd\over{\longrightarrow}& p_\mu
p^\mu=-m^{2} &\hbox{Poincar\'e-mass shell}\quad\cr
u\cdot u=1\ \ \ \ \ &\longrightarrow&
g_{\mu\nu}(x)a^\mu a^\nu=1\ \ \ \ \, & \buildrel
\kzsd\over{\longrightarrow}& a_\mu a^\mu=1\ \ \ \ \,
&\hbox{}\cr
v\cdot v=1\ \ \ \ \ &\longrightarrow&
g_{\mu\nu}(x)b^\mu b^\nu=1\ \ \ \ \,& \buildrel
\kzsd\over{\longrightarrow}& b_\mu b^\mu=1\ \ \ \ \,
&\hbox{}\cr
t\cdot t=m^2s^2&\longrightarrow&
g_{\mu\nu}(x)s^\mu s^\nu=m^2s^2 & \buildrel
\kzsd\over{\longrightarrow}& s_\mu s^\mu=m^2s^2
&\hbox{Pauli-Lubanski cdt.}\quad\cr
q\cdot
t=0\ \ \ \ \ &\longrightarrow& g_{\mu\nu}(x)p^\mu
s^\nu=0\ \ \ \ \, & \buildrel \kzsd\over{\longrightarrow}& p_\mu
s^\mu=0\ \ \ \ \ &\hbox{orthogonality cdts.}\ \quad\cr
y^5q^0-y^0q^5>0&\longrightarrow& p^0>0\quad(3.24a)\quad \
& \buildrel \kzsd\over{\longrightarrow}&p^0>0\qquad\quad
&\hbox{positive energy}.\,\qquad\ \cr
}$$
\medskip
\noindent Here the third column concerns the
Poincar\'e-invariant theory.  The scalar product appearing there
is the one associated to the Minkowski flat metric, which is the
zero curvature of $g$ in (2.3).  In the next to last row we
displayed only one example of the pseudo-orthogonality
relations.  The constraint (3.1g) becomes, when
\kz\ , $\epsilon_{\mu\nu\lambda\delta}p^\mu a^\nu b^\lambda
s^\delta=m^2s$.  Here $\epsilon_{\mu\nu\lambda\delta}$ is the
completely skew-symmetric on the Minkowski spacetime.  This
equation is also valid on $\e$ with its new coordinates,
$\epsilon_{\mu\nu\lambda\delta}$ will be then the completely
skew-symmetric tensor on $M_\k$ ($\det g=-1$).

We show now how we can recover in the zero curvature limit
the evolution space used by Souriau [So] for the case of a
mass $m$ and spin $s$ free particle on $M_\k$.  Actually, the
latter appears as the \kz\ limit of $\f$ (3.18). In fact, translating the
constraints defining $\f$, in the
$x$-coordinates and using the limits displayed above one can
easily see that they become in the \kz\ limit those used by
Souriau [So].  In order to apply the same procedure to $\omf$,
we first rewrite it in the $x$-coordinates,
$$\omf=dx\wedge
dp-{1\over2m^2}\,\epsilon_{\mu\nu\rho\sigma}\,p^\mu
s^\nu\,\Big\lbrack\k^2\,dx^\rho\wedge dx^\sigma+{1\over
m^2}\,Dp^\rho\wedge Dp^\sigma-{1\over m^2s^2}\,Ds^\rho\wedge
Ds^\sigma\Big\rbrack.\eqno(3.26)$$ Here
$\epsilon_{\mu\nu\rho\sigma}$ is the completely
skew-symmetric tensor on $M_\k$, $dx\wedge
dp=dx^\mu\wedge dp_\mu$ with $p^\mu=g^{\mu\nu}\,p_\nu$
and $Dp$ is the covariant differential of $p$, i.e.
$$Dp_\mu=dp_\mu-\Gamma^\rho_{\mu\sigma}\,p_\rho
dx^\sigma,\qquad\mu, \nu, \rho\ \hbox{et}\ \sigma\in\{0, 1, 2,
3\};\eqno(3.27)$$ $\Gamma^\mu_{\rho\sigma}$ is the affine
connection  corresponding to the AdS metric (2.3).  Then we
take the \kz\ limit.  We obtain,
$$\omf^0=\lk \omf=dx\wedge
dp-{1\over2m^2}\,\epsilon_{\mu\nu\rho\sigma}\,p^\mu
s^\nu\,\Big\lbrack{1\over m^2}\,dp^\rho\wedge dp^\sigma-{1\over
m^2s^2}\,ds^\rho\wedge ds^\sigma\Big\rbrack.\eqno(3.28)$$
This is exactly the presymplectic form Souriau used in his
work [So].  Thus, $\f$ can be considered as the AdS deformation
of the evolution space describing the theory of a free massive and
spinning particle on the Minkowski spacetime.

Finally, note that all the limits we evaluated in this subsection confirm
the physical interpretation we gave to the AdS
quantities previously introduced.

\bigskip\noindent{\subsect 3.3. The coadjoint orbit contraction}
\medskip
\noindent In the previous subsection we showed that  the
presymplectic manifold $\f$ is the AdS deformation of the
evolution space used by Souriau in its description of a massive and
spinning free particle on Minkowski spacetime. Here we
investigate the zero curvature behaviour of the phase space $\s$.
In order to stay close to the group theoretical meaning of
contraction, this point is discussed in the language of coadjoint
orbits.  In fact $\s$ is diffeomorphic to a coadjoint orbit
$\orb$ of $\so$ [Ko] [SW].

We first start by identifying
$\orb$.  This is achieved through the moment map in the
following way.  Usually the momentum map is defined as a map
from the phase space into the dual of the Lie algebra.  However
in the present case it can be defined as a map from the evolution
space into $\ssoo$.  The two constructions are equivalent as
shown below.  In fact, let the map $L$ be defined as follows:
$$L:\e\ni\ddbv\,\longmapsto
L(\ddbv)\in\ssoo\eqno(3.29a)$$ such that,
$$\langle
L(\ddbv)\,, e_{\alpha\beta}\rangle=
\X\,\rfloor\te\equiv\L(\ddbv),\qquad\ab.\eqno(3.29b)$$
The symbol $\langle\ ,\ \rangle$ denotes the duality
$\soo$-$\ssoo$ and the $e_{\alpha\beta}$'s are the basis
elements of the abstract $\soo$ algebra (see (2.4)).  The $\X$'s are the
fundamental vector fields associated to the (left) action of $\so$
on $\e$.  The $\L$'s introduced in (3.29b) are the {\it classical
observables\/} associated to the action of $\so$ on $\e$.  Their
explicit form is easily obtained:
$$\eqalignno{\L
&=y_\alpha q_\beta-y_\beta
q_\alpha+s\,(u_\alpha v_\beta-u_\beta v_\alpha)&(3.30a)\cr
&=y_\alpha q_\beta-y_\beta q_\alpha+{\k\over m^2}
\epsilon_{\alpha\beta\gamma\rho\sigma}y^\gamma q^\rho t^\sigma,
\qquad\forall\alpha, \beta
\in \{5, 0, 1, 2, 3\}.&(3.30b)\cr}$$
The second expression is a consequence of the orientation
condition in (3.1g).  These observables are constants of the
motion, i.e. $d\L\,\rfloor\ke=0$,
where we recall that $\ke(\ddbv)\in\ker_{_\dbv}\ome, \forall
\ddbv\in\e$ is given in (3.15).  As a result, the usual moment map
$\widetilde L: \s\longrightarrow\ssoo$ is obtained through the symplectic
reduction $\pi: \e\longrightarrow\s$, i.e. $L=\widetilde L\circ\pi$.
The associated classical observables $\wL\in C^\infty(\s)$ are
then given by $\wL(\ww)=\L(\ddbv)$, where $\ww=\pi(\ddbv)$.
The $\wL$'s realize $\soo$ through the Poisson bracket defined
by the symplectic form $\os$.  Since $\so$ is a simple Lie group, its action
on $\s$ is strongly Hamiltonian, the momentum map is
uniquely defined and also equivariant [LM].  The image of $\s$ under
$\widetilde L$ is then an orbit in $\ssoo$.  This is the coadjoint orbit we
shall denote $\orb$.  Note that for
$\wo\in\e$ given in (3.2)
$$L(\wo)=\mk\,\theta^{50}+s\,\theta^{12};\eqno(3.31)$$
hence $\orb$ passes through $\te$.  This is not a coincidence, it
is just a consequence of the general theory [SW].  The orbit
$\orb$, can be realized through constraint equations in
$V\equiv \R^{10}$ (the vector space underlying $\soo$ or
$\ssoo$).  These equations are provided by the Casimir
invariants of $\soo$.  Actually, a straightforward computation
gives the two identities,
$$\demi\,\L L^{\alpha\beta}={m^2\over\k^2}+s^2,
\eqno(3.32a)$$
$$\Pi_\alpha\Pi^\alpha={m^2s^2\over\k^2},\eqno(3.32b)$$
where $\Pi_\alpha ={1\over 8}
\epsilon_\alpha{}^{\mu\nu\rho\sigma}
L_{\mu\nu} L_{\rho\sigma}$.  Notice that these equations do
not define a connected submanifold of $V$.  However, the
connected component $\orb$ is uniquely specified by imposing
that it passes through $\te\in V$.  Two remarks are now in order:

\item{\bf 1.\enskip}{The invariants above have the same value
for two distinct physical sytems $(m, s, \k)$ and $(m', s', \k')$,
such that $\mk=s'$ and ${m'\over\k'}=s$,
however the corresponding $\te$ belong then
to two distinct orbits.  This phenomenon can be viewed as a consequence of
a classical counterpart of the Weyl symmetry [GH].  In order to restrict
ourselves to only one of the two types of orbits, we will from now on
consider as physical only triplets \msk such that $\mk>s>0$.  At first sight
this condition seems to be weakly justified.  Its physical origin will
be discussed in connection with previous works  in the next
subsection.}

\item{\bf 2.\enskip}{The same orbit, $\orb$, is associated to
two distinct physical systems: $(m, s, \k)$ and $(m', s', \k')$, such
that $\mk={m'\over\k'}$ and $s=s'$.  Hence, to the contrary of
the Poincar\'e group, the correspondence between physical
systems and elementary systems (i.e. coadjoint orbits) is not
one-to-one.  Once again, this point will be discussed in the next
subsection.  In fact we will show that 1 and 2 are related.

\noindent We come now to the contraction of the orbit $\orb$.  To
this end we will use a sequence of $\k$-dependent
transformations, such that when $\k$ tends to zero the
transformed orbit (which is no longer an orbit) tends to a
Poincar\'e-coadjoint orbit (see [Do]).  Recall that the contraction map
$\phi_\k$ introduced in (2.5) allowed us to reach $p(3,1)$
starting from $\soo$.  Using $\phi_\k$, we define in a natural way the
following family of maps:
$$\widetilde L^\k=\phi^*_\k\circ \widetilde L:
\s\,\longrightarrow\, V^*.\eqno(3.33)$$
In analogy with (3.29), we shall write,
$$\widetilde L^\k(\ww)=\demi
\wL^\k(\ww)\,\vartheta^{\alpha\beta}.\eqno(3.34)$$
where $\{\vartheta^{\alpha\beta}\}$ is the basis of $\ssoo$ dual
to $\{e_{\alpha\beta}\}$.  More precisely, the $\wL^\k$'s are related to the
$\wL$'s in the same way the $e^\k_{\alpha\beta}$'s are related to
$e_{\alpha\beta}$'s in (2.5).  In order to evaluate the \kz\ limit, we first
express the $\wL^\k$'s in the contraction adapted coordinates.  Concretely,
using (3.30), the results of subsection 3.2 and the explicit symplectic
reduction of subsection 3.1, one finds
$$\eqalignno{\widetilde L^\k_{50}&=-p_0+{\k^2\over
m^2}\sum_{i,j,k=1, 2,
3}\epsilon_{ijk}\,(x^ip_js_k)&(3.35a)\cr
\widetilde
L^\k_{i5}&=\k Y\,p_i+{\k^2\over m^2\,(\ky)}\sum_{j,k=1, 2,
3}\epsilon_{ijk}\,\Big\lbrack \,x^j \,(p_0s_k-s_0p_k)
\Big\rbrack,&(3.35b)\cr
\widetilde
L^\k_{0i}&=-{p_0\over(\ky)}\,x^i+{\ky\over m^2}\sum_{j,k=1, 2,
3}\epsilon_{ijk}\,p_js_k,&(3.35c)\cr
\widetilde
L^\k_{ij}&=(x^ip_j-x^jp_i)-{1\over m^2}\sum_{k=1, 2,
3}\epsilon_{ijk}\,(p_0s_k-s_0p_k).&(3.35d)\cr}$$
Here $i, j \in \{1, 2, 3\}$ and $\epsilon_{ijk}$ is the completely
skew-symmetric tensor of $\R^3$.  When \kz\ they become
respectively, $H$, $P_i$, $K_i$ and $J_i$ given by,
$$\eqalignno{H&= p^0=\sqrt{(\vec p)^2+m^2},&(3.36a)\cr
\ovp&=\vec p,&(3.36b)\cr
\ovk&=H\vec x+{\vec p\times \vec
s\over m^2},&(3.36c)\cr
\ovj&=\vec x\times\vec p+{p^0\vec
s-s^0\vec p\over m^2} \cdotp &(3.36d)\cr}$$ One can easily
recognize these observables as those associated to the Poincar\'e
group, even if they are written in an unusual representation.  In
fact the position $\vec x$ arising here, is naturally the world line
position ($x^\mu$ is a four vector) but not the canonical position
usually appearing in the literature [SM] [He] [M].  The usual
realization is recovered through the use of the following
transformation, $$\ovx=\vec
x+{\vec p\times\vec s\over m^2(p^0+m)},\eqno(3.37a)$$
$$\ovs={\vec
s\over m}-{s^0\vec p\over m(p^0+m)};\eqno(3.37b)$$
$\ovx$ is the
canonical position, $\ovs$ lies on the sphere of radius $s$,
$\ovs\cdot \ovs=s^2$ and $s^\mu$ is, according to subsection
3.2, the Pauli-Lubanski four-vector.  In terms of $\ovx$ and
$\ovs$, $\ovk$ and $\ovj$ are then
$$\eqalignno{\ovk&=p^0\ovx+{\vec p\,\times\ovs\over
p^0+m},&(3.38a)\cr
\ovj&=\ovx\times\,\vec
p\,+\,\ovs.&(3.38b)\cr}$$
Equations (3.36a-b) and (3.38a-b) are the usual Poincar\'e
observables associated to a mass $m$ and spin $s$ free particle
on Minkowski spacetime.  The casimir invariants that allow the
identification of the $\ppfl$-coadjoint orbit obtained here arise
also as a \kz\ limit of (3.32a-b).  Actually, using (3.36a-d) we
show that,
$${\k^2\over 2}\,\wL
\widetilde L^{\alpha\beta}=m^2+\k^2s^2 \quad\buildrel
\kzsd\over{\longrightarrow}\quad
H^2-\ovp\cdot\ovp=m^2,\eqno(3.39)$$
$$\lk\widetilde\Pi_5^\k=0,
\qquad\lk\widetilde\Pi_0^\k=\ovp\cdot\ovj=s^0,\qquad
\lk\overrightarrow{\widetilde\Pi^\k}=H\ovj-\ovk\times\ovp=\vec
s.\eqno(3.40)$$ The tilde in (3.40) means that we consider (3.32b)
with the $\wL$'s instead of the $\L$'s.  Equations in (3.40) are
the well known expressions defining the Pauli-Lubanski
four-vector $s^\mu$ in terms of the Poincar\'e generators [SM], in
fact $$-(s^0)^2+(\vec s)^2=m^2s^2.\eqno(3.41)$$
We conclude that in the zero curvature limit, the surfaces
$\widetilde L^\k(\s)\subset V^*$ tend to the $\ppfl$-orbit $\opoin$,
corresponding to a mass $m$ and spin $s$ free particle on Minkowski
spacetime, which passes through $m\vartheta^{50}+s\vartheta^{12}\in
V^*$.  In other words, $\widetilde L^0$ is a $\ppfl$-equivariant moment
map.

\bigskip

\noindent{\subsect 3.4. Discussion}
\medskip
\noindent In subsection 3.3 we restricted the notion of a physical
system to those triplets \msk satisfying $\mk>s>0$.  We imposed
this condition in order to associate to physical systems a unique
type of coadjoint orbits, namely those passing through
$a\,\vartheta^{50}+b\,\vartheta^{12}$ such that $a>b$.  One can
end up with this condition from other considerations.  In fact, as
we noticed in remark 2 of subsection 3.3, the same coadjoint orbit
can be associated to different physical systems.  This is clearly
due to the fact that a physical system is specified by three
parameters while  the corresponding coadjoint orbit is specified
by two parameters.  In order to compare this situation with the
one appearing in the flat case one must first fix the
AdS spacetime, i.e. fix the curvature $\k$,  and then investigate
the one-to-one character of the correspondence.  Doing so, one
has to solve the system of equations arising from (3.32a-b),
where $m$ and $s$ are the unknowns for given values of
${m^2\over\k^2}+s^2$ and ${m^2s^2\over\k^2}$.  This
problem has two possible solutions satisfying either $\mk>s$ or
$\mk<s$.  By considering only one of the latter as physically
realizable, we obtain an AdS analog of the one-to-one
correspondence (physical system $\leftrightarrow$ coadjoint
orbit) occuring in the case of the Poincar\'e group.  Here we
choose $\mk>s$ as the physical condition.  This choice is encouraged by
concording arguments used by different authors, see
for instance [Di] [Ku] [Wo].

A purely classical argument can be found in [Di].  In fact, there
the author considers a free extended object on $M_\k$.  The
condition $\mk>s$ appears as a reasonable physical one, since
$\mk\leq s$ implies that the extended object spin with a speed
larger than that of light and has dimensions greater than the
radius $\k^{-1}$ of the universe!

The arguments found in [Ku] and [Wo] can be applied to our
present work since they arise from the study of the classical
dynamics of an elementary particle in general relativity.  The first
author imposes the general condition
$(spin)\times(curvature)<(mass)$ in order to avoid the region
where the equality holds, since then the dimension of the kernel
of the presymplectic form becomes larger (as it happened here,
see (ii) in subsection 3.1).  He shows also that this condition is
valid even in the extreme situation of an electron near the horizon
of a Schwarzschild black hole.  The second author used the
following argument: for spin values of the order of $\hbar$ the
condition $\mk>s$ (with $\hbar$ and $c$ no longer equal to $1$)
is violated by particles having a Compton wavelength ${h\over
mc}$ of the same order or larger than the radius of the universe.
The one particle theory fails then, since  in that case  the
gravitational force is strong enough to induce the creation of
pairs of particles.

Finally, let us mention that this constraint fits very well with the
fact noticed in (ii) (subsection 3.1).  In fact, as for the Poincar\'e
case where $m>0$ specifies a massive elementary system and
the lower limit on $m$ a massless one, the lower limit $\mk=s$
here corresponds to a massless AdS elementary system [E2].
Observe also that when \kz\ the condition $m>\k s$ becomes
simply $m>0$.  Moreover, it will appear in section 4 that $\mk>s$ is a
necessary condition for the unitarity of the representation obtained.

\bigskip

\noindent {\sect 4. The quantum theory}
\medskip \noindent
The aim of this section is to construct the quantum theory of a
mass $m$ and spin $s$ free particle on AdS spacetime of
curvature $\k>0$.  To this end we shall quantize the classical
theory described in the previous section.  More precisely, we will
use geometric quantization techniques, which exploit in an
efficient way the geometric constructions of section 3.  It is
well known the quantum theory we are looking for is described
by a unitary irreducible representation (UIR) of $\so$ [Fr2]
[Wi1].  This representation can be obtained by geometric quantization,
known also as the orbit method of Kirillov [Ki1]
[Ki2].  This particular method has the advantage of allowing us to identify
the physical interpretation of the quantities appearing in the quantum
theory, such as the quantum states and the observables, since it is based
on the classical theory, where the physical interpretations are already
established.  Note that the spacetime realization of the representation do
not have this feature [Fr2].

The geometric quantization proceeds in two steps.  First, one follows a
prequantization procedure that identifies a unitary but reducible
representation of $\so$ (section 4.1).  Then one uses
polarization conditions that select an irreducible
subrepresentationb (section 4.2).  Since this programme has already been
carried out in [DBE] for the $1+1$ dimensional case, we will
omit here unnecessary details.  For the general theory of geometric
quantization we refer to [Ko] [SW] and [Wo].

Finally let us mention that the results of section 3 greatly
simplify the computations.  In fact, we shall base our
construction on $\e$, i.e. all relevant quantities will be defined
on $\e$, keeping in mind that the quantities on the phase space
$\s$ can be derived making use of the symplectic reduction of
subsection 3.1.

\bigskip\noindent{\subsect 4.1. Prequantization}
\medskip
\noindent
The prequantum Hilbert space $\h$, when it exists, is generally
defined as the space of square integrable sections of a Hermitian
line bundle-with-connection over the phase space, such that the
symplectic form of the latter is the curvature of that connection.
The existence of $\h$ requires that the symplectic form satisfies
an integrability condition [Wo].

The identification of $\h$ in the present case is greatly simplified
because of the principal bundle structure $\pi:
\e\cong\so\longrightarrow\so/SO(2)\times SO(2)\cong\s$,
explicitly realized in section 3.  In fact, $\h$ consists then of
functions $\psi\in L^2(\e, d\mu^{m,s}_\k)$, satisfying the
condition:
$$(K\psi)(\ddbv)=i(K\,\rfloor\te)\psi(\ddbv),\eqno(4.1)$$
for all vector field $K$ on $\e$,  such that $K(\ddbv) \in
\ker_\dbv\ome\ \forall\, \ddbv\in \e$ [Wo].  Note that
$d\mu^{m,s}_\k$ is the invariant measure on $\e$, obtained from
the left Haar measure on $\so\cong\e$.  Recalling from
(3.15) that $\ker\ome$ is generated by $Y_{50}$ and
$Y_{12}$  given in (3.5a) we can explicitly write,
$$\h=\left\{\psi:\e\longrightarrow\C\
\Big\vert\ \int_{\e}\vert\psi\vert^2d\mu^{m,s}_\k<\infty,\enskip
Y_{50}\psi=i\mk\,\psi\enskip\hbox{and}\enskip
Y_{12}\psi=is\,\psi\right\}.\eqno(4.2)$$
The integrability condition mentioned in the first paragraph above
appears here as a condition
for the integrability of the equations in
(4.2) to the global group action .  Hence, since $Y_{50}$ and
$Y_{12}$ generate a compact subgoup of $\so$, this implies
that $\mk$ and $s$ must be integers.  One can also view this
condition in terms of the integrability of $\te$ given in (3.6).

Clearly the quantum theory that will arise from the present
quantization will only describe integer spin elementary systems.
In order to take also into account
the half integer spin particles one should, from the begining,
consider as the symmetry group of the theory $Sp(4, \R)$
instead of $\so$.  The former is the double covering of the latter
(see for instance [BEGG]).  This can also be achieved by
considering the universal covering group of $\so$ as the
symmetry group, in this case $\mk$ will no longer be an integer,
it will then take its values in $\R^*_+$.

The Hilbert space $\h$ carries a unitary representation of $\so$.
In fact, since $\e\cong\so$, there exists a natural action of $\so$ in
$L^2(\e, d\mu^{m,s}_\k)$.  This yields the left regular
representation of $\so$ denoted $U(\Lambda)$ and given by,
$$\big(U(\Lambda)\psi\big)(\ddbv)=\psi(\Lambda^{-1}\cdot
\ddbv),\quad\hbox{where}\quad(\Lambda^{-1}\cdot
\ddbv)^\mu=(\Lambda^{-1})^\mu{}_\nu\ddbv^\nu.\eqno(4.3)$$
When restricted to $\h$ it provides us with a unitary
representation of $\so$ denoted by $(\h, U)$.  In the language of
induced representations $(\h, U)$ is a representation of $\so$
induced from the unitary character $\exp i(\mk\tau+s\tau')$ of
the subgroup $SO(2)\times SO(2)\subset\so$.

 From (4.3) we can obtain the expression of the (pre)quantum
operators $\hL$, i.e. the quantum analogs of the $\L$'s given in
(3.30).  Actually,
$$\hL\equiv i{d\over d\tau}\Big(U(\exp\tau
e_{\alpha\beta})\psi\Big)(\ddbv)
\Big\vert_{\tau=0}=-i(\X\psi)(\ddbv).\eqno(4.4)$$
\nobreak The $\hL$'s are then nothing but $(-i)$ times the fundamental
vector fields $\X$ associated to the action of $\so$ on $\e$.

\bigskip\noindent{\subsect 4.2. Polarization}
\medskip
\noindent
Exploiting once again the principal bundle structure $\pi:
\e\longrightarrow\s$, we
can use an algebraic characterization of the polarization on $\s$ in
order to concretely evaluate it [Ra1] [Re] [Wo].  Actually this
characterization determines a prepolarization on $\e$ the
projection of which on $\s$ produces an invariant polarization.  Since we
are interested in evaluating the quantum theory at the level of
$\e$, we will only need the prepolarization
[Wo].

The latter is a subalgebra $h$ of $\soc$,  the complexified
$\soo$, which satisfies the following conditions:
\itemitem{{\bf (i)}\quad}$Y_{50}$ and $Y_{12}\in h$.
\itemitem{{\bf (ii)}\quad}$\dim_{\cp}
h=\demi\left(\dim\soo+\dim\ker\ome\right)\enskip\Longrightarrow
dim_{\cp}h=6$ since $\dim\ker\ome=2$.
\itemitem{{\bf (iii)}\quad}$\te\big(\lbrack Y,
Y'\rbrack\big)=0,\quad\forall\ Y, Y' \in h$.   \itemitem{{\bf
(iv)}\quad}$h+\bar h$ is a subalgebra of $\soc$.

\noindent  The prepolarization projects
on $\s$ to a {\it K\"ahler\/} (resp. {\it positive\/}) polarization if
$h\cap \bar h=\{Y_{50},Y_{12}\}$ (resp. $i\te\big(\lbrack Z, \overline {\!
Z}\rbrack\big)\geq0,\ \forall\ Z\in h$).

It is easy to check that the following subalgebra $h$ of $\soc$,
$$h={\rm span}\,\{Y_{50}, Y_{12}, Z_1, Z_2, Z_3,
\Xi\},\eqno(4.5a)$$ where
$$Z_i=Y_{0i}+iY_{i5}, \quad \iu\quad\hbox{and}\quad
\Xi=Y_{23}+iY_{31},\eqno(4.5b)$$
is a prepolarization on $\e$.  Moreover, its projection on $\s$ is a
K\"ahler, positive and invariant polarization.  Notice that $\Xi$
in (4.5b) is, at least algebraically, the usual K\"ahler
polarization of a sphere. The latter is in our case the
homogeneous space for the subgroup $SO(3)\subset\so$
generated by the $Y_{ij}, i, j \in \{1, 2, 3\}$.  Hence, $\Xi$
characterizes the spin contribution in $\e$.

We are now able to select in $(\h, U)$ the unitary irreducible
subrepresentation $\u$ we are looking for [Wo].   This is just the
restriction of $U$ to $\hmsk$, where
$$\hmsk=\left\{\psi\in\h\ \big\vert\
\overline{\!Z}_i\psi=0,\enskip\iu\quad\hbox{and}\quad
\overline{\Xi}\psi=0\right\}.\eqno(4.6)$$
The way we obtained the UIR $(\hmsk, \u)$
is called in mathematics litterature a holomorphic
induction [Hu] [Sc].  It produces discrete series representations of
noncompact semi-simple Lie groups.  The one we obtained here is
the quantization of the orbit $\orb$ of section 3.  It is the
discrete series representation of $\so$ characterized by the highest
weight $(\mk, s)$ associated to the Cartan subalgebra generated by
$e_{50}$ and $e_{12}$.  Moreover, it is also known that
the unitarity of $\u$ requires the necessary condition $\mk>s$
[Fr1] [Ev].  Notice that the physical constraint $\mk>s$ we
imposed in the classical theory (see section 3.4) finds in the
quantum paradigm an interpretation in terms of unitarity.  A
more restrictive necessary and sufficient unitarity condition can
be found in [Fr2] [FH].  It can be recovered in the present
construction using the same method we used in [DBE].

The quantum states are represented by well defined wave
functions belonging to $\hmsk$.  The physical interpretation of
their modulus as probability densities on $\s$ is also well
defined.  Notice that this very important quantum property
is inherent to the phase space representation, it lacks in the other
known representations (spacetime or momentum space
representations), and it clearly arises from the square
integrability of the representation $(\hmsk, \u)$.  This property constitutes
the
basic ingredient necessary for the definition of the notion of
optimal localization that will be given in the next section.

Finally, let us write the complex
coordinates induced by the K\"ahler polarization on
$\e$.   These are $z$ and $\xi\in(\C^5, \eta)$ given by,
$$z=\k\,y-im^{-1}\,q\quad\hbox{and}\quad\xi=u-iv.\eqno(4.7)$$
Here $y$, $q$, $u$ and $v$ are the vectors introduced in section
3.1.  The presymplectic form $\ome$ in (3.10) becomes in terms
of these new coordinates, $$\ome=i\mk\, dz\wedge d\bar z+ is\,
d\xi\wedge d\bar\xi. \eqno(4.8).$$
All the quantities introduced up to now can be reexpressed in
terms of $z$ and $\xi$.  Moreover, the phase space $\s$ can be viewed as
a symplectic reduction of $(T\C^5, \eta)$ equipped with the canonical
symplectic form $\ome$ given above.

\bigskip

\noindent {\sect 5.  Optimal localization and its
zero curvature limit}
\medskip \noindent
In section 3 we constructed the classical theory and we
investigated its zero curvature limit.  More precisely, we
evaluated the \kz\ limit of the orbit $\orb$ and we found that it
gives rise to a $\ppfl$-coadjoint orbit $\opoin$.  In section 4 we
quantized $\orb$.  In order to complete the picture we want now
to investigate the zero curvature limit of this quantum theory.
We start in this section by studying the \kz\ behaviour of a particular
family of quantum states $\fww\in\hmsk$ indexed by the points
$\ww\in\s$.  We shall identify the $\fww$ through the notion of optimal
localization on $\s$ that we now introduce.  First, recall that since $\s$ is
an $8$-dimensional $\so$-homogeneous space, a point
$\ww\in\s$ is completely specified by giving the values of $L_{31}$,
$L_{23}$, $L_{0i}$ and $L_{i5}$  $\iu$ at $\ww$.  This suggests the
following definition.  We shall say a state $\varphi\in\hmsk$ is localized
at the point $\ww$ in phase space, if the quantum expectation values of
the eight observables $\hL$'s with $(\alpha, \beta)=(0, i), (i, 5), (3, 1)$ or
$(2, 3)$ equal the corresponding classical values, i.e.
$$\langle\varphi\mid\hL\mid\varphi\rangle=\L(\ddbv),\eqno(5.1)$$
when $\ww=\pi(\ddbv)$.  It is not hard to see that these eight conditions do
not specify $\varphi$ uniquely.  If on the other hand we require (5.1) to
hold for all ten group generators, then $\varphi$ is uniquely determined (up
to a phase) and we write $\fww$ for the solution.  For reasons explained
shortly, we shall say $\fww$ is the state {\it optimally\/} localized at
$\ww$.  We now compute the $\fww$ explicitly.  First, consider the state
$\fo\equiv\varphi_{_{\tilde\dbv_{(0)}}}$ with $\wo\in\e$ defined in
(3.2). Equation (5.1) yields,
$$\langle\fo\mid\hat
L_{50}\mid\fo\rangle=\mk,\quad
\langle\fo\mid\hat
L_{12}\mid\fo\rangle=s\quad\hbox{and}
\quad\langle\fo\mid\hL\mid\fo\rangle=0\enskip\hbox{otherwise}.
\eqno(5.2)$$
Hence $\fo$ is the highest weight
vector in $\hmsk$.   It is then immediately clear that,
$$\fww=\u\big(\Lambda(\ddbv)\big)\,\varphi_0, \eqno(5.3)$$
where $\Lambda(\ddbv)$ is the $\so$ element given in (3.3).  We conclude
that the states $\{\fww \mid \ww\in\s\}$ belong to the orbit $\of$ of
the action of $\u$ on the highest weight vector $\fo$.

The optimal character of the localization of the states
$\fww\in\of$ arises from a known property of these states.  In
fact, they minimize the dispersion relations associated to the
Casimir invariants of $(\hmsk, \u)$ [De] [DF] [Pe].
The explicit derivation of these results will not be given
here since it is a straightforward generalization of
those obtained in [DBE].  According to Perelomov [Pe] the
states in $\of$ are called generalized coherent states.  Note that in the
subsequent we will denote $\fww$ equivalently by $\fw$.

Let us now evaluate $\fw\in\of$.  To this end we first need to
determine $\fo$.  This is actually realized through solving the
following system of equations, which arise from (4.2), (4.6) and
(5.2),
$$\openup
1.5mm\eqalignno{
Y_{50}\,\fo=i\mk\,\fo&\,\Longrightarrow\,(\bar
z\cdot{\partial\over\partial\bar z}-z\cdot{\partial\over\partial
z})\,\fo=\mk\,\fo,&(5.4a)\cr
Y_{12}\,\fo=is\,\fo&\,\Longrightarrow\, (\bar
\xi\cdot{\partial\over\partial\bar\xi}-\xi\cdot{\partial\over\partial
\xi})\,\fo=-s\,\fo,&(5.4b)\cr
\overline{\!Z}_+\,\fo=0&\,\Longrightarrow\,(\bar
\xi\cdot{\partial\over\partial\bar z}+z\cdot{\partial\over
\partial\xi})\,\fo=0,&(5.4c)\cr
\overline{\!Z}_-\,\fo=0&\,\Longrightarrow\,(\xi\cdot{\partial\over\partial\bar
z}+z\cdot{\partial\over \partial\bar\xi})\,\fo=0,&(5.4d)\cr
\overline{\!Z}_3\,\fo=0&\,\Longrightarrow\,
t\cdot{\partial\over\partial\bar z}\,\fo=0,&(5.4e)\cr
\overline\Xi\,\fo=0&\,\Longrightarrow\,
t\cdot{\partial\over\partial\bar \xi}\,\fo=0,&(5.4f)\cr}$$ $$\hat
L_{50}\,\fo=\mk\,\fo
\Longrightarrow\lbrack(z_5{\partial\over\partial z^0}-
z_0{\partial\over\partial z^5})+(\bar z_5{\partial\over\partial
\bar z^0}- \bar z_0{\partial\over\partial \bar
z^5})+(z\rightarrow\xi)\rbrack\,\fo=i\mk\,\fo,\eqno(5.4g)$$
$$\hat L_{12}\,\fo=s\,\fo
\Longrightarrow\lbrack(z_1{\partial\over\partial z^2}-
z_2{\partial\over\partial z^1})+(\bar z_1{\partial\over\partial
\bar z^2}- \bar z_2{\partial\over\partial \bar
z^1})+(z\rightarrow\xi)\rbrack\,\fo=is\,\fo.\eqno(5.4h)$$
Some notational precisions are in order.
Actually, $\overline{\!Z}_+={i\over
2}(\overline{\!Z}_1+i\,\overline{\!Z}_2)$ and $\overline{\!Z}_-={i\over
2}(\overline{\!Z}_1-i\,\overline{\!Z}_2)$, for $Z_1$ and $Z_2$
given in (4.5b).  The vector fields in (5.4) do not contain any
derivatives with respect to $t^\alpha$.  This is due to a
transformation we made in order to express the equations above
only in terms of the complex coordinates $z$ and $\xi$.  Thus,
$\fo(z, \bar z, \xi, \bar\xi)\equiv\psi_0(z, \bar z, \xi, \bar\xi, t(z,
\bar z, \xi, \bar\xi))$, where $\psi_0$ is the vector that
originally appears in (4.6) and $t^\alpha=-{ms\over
4}\,\epsilon^\alpha{}_{\beta\gamma\rho\sigma}z^\beta \bar
z^\gamma\xi^\rho\bar\xi^\sigma$ (see (3.1g)).

It is easy to check that the solution of the above system of
equations is given by,
$$\fo(z,
\xi)=N\Big(\bzo\cdot z\Big)^{-\mk-s}\Big\lbrack(\bzo\cdot z)(\bxo\cdot
\xi)-(\bzo\cdot \xi)(\bxo\cdot z)
\Big\rbrack^s;\eqno(5.5)$$
$N$ is a normalization constant and
$\zo=\k\yo-im^{-1}\qo$ and
$\xo=\uo-i\vo$.  This solution is well defined on $\e$.  In fact,
one easily verifies that $\bzo\cdot z=z_5+i\,z_0$ vanishes only
when $y^5q^0-y^0q^5<0$.  Because of equation (3.1h) this can
never happen for points of $\e$.
Using (5.3), $\varphi_{\dbv'}\in\of$ is obtained as follows,
$$\openup
1.5mm\eqalignno{\varphi_{\dbv'}(z,\xi)
&=\u\big(\Lambda(\ddbv')\big)\fo(z, \xi)\cr
&=\fo\big(\Lambda^{-1}(\ddbv') z, \Lambda^{-1}(\ddbv')
\xi)\big)\cr
&=N\left(\bzo\cdot
\Lambda^{-1}z\right)^{-\mk-s}\left[(\bzo\cdot
\Lambda^{-1}z)(\bxo\cdot \Lambda^{-1}\xi)-
(\bzo\cdot \Lambda^{-1}\xi) (\bxo\cdot
\Lambda^{-1}z) \right]^s\cr
&=N(\Lambda\bzo\cdot
z)^{-\mk-s}\,\lbrack(\Lambda\bzo\cdot z)(\Lambda\bxo\cdot
\xi)-(\Lambda\bzo\cdot \xi)(\Lambda\bxo\cdot z) \rbrack^s\cr
&=N(\bar z'\cdot
z)^{-\mk-s}\,\lbrack(\bar z'\cdot z)(\bar \xi'\cdot \xi)-(\bar
z'\cdot \xi)(\bar\xi'\cdot z) \rbrack^s.&(5.6)\cr}$$
Here we used (4.3) and also the fact that
$\Lambda(\ddbv')\wo=\ddbv'$ (see (3.3)).  Notice that we can
equally well write $\varphi_{\dbv'}(z,\xi)$ as
$\varphi_{(z',\xi')}(z,\xi)$.  The normalization is fixed by
imposing that $\varphi_{(z,\xi)}(z,\xi)=1$.  This gives
$N=(-2)^{\mk}(2)^{-s}$.  Finally the optimally localized state
at $(z', \xi')\in\e$ is given by,
$$\varphi_{z', \xi'}(z, \xi)=(-2)^{\mk}(2)^{-s} \Big(\bar z'\cdot
z\Big)^{-\mk-s}\Big\lbrack(\bar z'\cdot z)(\bar \xi'\cdot \xi)-(\bar
z'\cdot \xi)(\bar\xi'\cdot z) \Big\rbrack^s.\eqno(5.7)$$
The optimal localization property can be read from (5.7).  In fact,
the modulus of $\varphi_{z', \xi'}(z, \xi)$ reaches its maximal
value only when $z=z'$ and $\xi=\xi'$.  Combining this with
the physical interpretation of the modulus of the states of
$\hmsk$ as probability densities on $\s$ one sees that
$\varphi_{z', \xi'}(z, \xi)$ is actually optimally localized at
$\ww'\in\s$.  If we consider, instead of the phase space
realization, the spacetime one we find that the state
corresponding to $\varphi_{z', \xi'}(z, \xi)$ is localized along
the timelike geodesic that arises from the projection on $M_\k$
of the leave of $\ker\ome$ passing through $\ddbv'\in\e$.  This
has been shown for the $1+1$ dimensional case in [DBEG].

Rewriting (5.7) in terms of the contraction adapted coordinates
of section 3.2, we are able to evaluate its zero curvature limit.
Using the same techniques as in [DBE], we obtain
$$\lk\left({m\over4\pi\k}\right)^{3\over2}\varphi_{z', \xi'}(z,
\xi)=m^2p^0\delta(\vec p-\vec
p\,')\,e^{-ip_\mu(x'^\mu-x^\mu)}\,\left({\bar\zeta'\cdot
\zeta\over2}\right)^s.\eqno(5.8)$$
Here, $\zeta_\mu=a_\mu-ib_\mu,
\ \mu\in\{0, 1, 2, 3\}$ and $(x, p, a, b, s)\in\ez$, $\ez$ being the
Lorentz bundle over Minkoski spacetime which is the  \kz\
limit of $\e$, see section 3.2.

The limiting state is clearly a distribution in the space of
$\C$-valued functions on $\ez$.  Moreover, notice that the
spin part $({\bar\zeta'\cdot
\zeta\over2})^s$ and the orbital part factorize separately. More
precisely, this state is perfectly localized in momentum space,
completely delocalized in spacetime though still optimally
localized in spin coordinates.  A further analysis of this result will be
given in the next section where we will show how the notion of optimal
localization is intimately related to the K\"ahler character of
the $\so$-invariant polarization.  We shall see that the zero curvature limit
of the latter produces a $\ppfl$-invariant polarization, which is no
longer K\"ahler, and relate this to the disappearance of the notion of phase
space localizatization.

\bigskip

\noindent {\sect 6. About the contraction of the discrete series}
\medskip
\noindent Exploiting the fact that the
construction of $(\hmsk, \u)$ in section 4 is $\k$-dependent, we
will investigate here the \kz\ limit of that construction.  Knowing
that the $\so$-coadjoint orbit $\orb$ becomes in the zero
curvature limit the $\ppfl$-coadjoint orbit $\opoin$ (see section
3.3), we expect that the \kz\ limit of the construction mentioned
above will produce the irreducible unitary representations of
$\ppfl$, obtained by quantization of $\opoin$.  Notice that the use of ideas
from geometric quantization to study the contraction of Lie groups
representations was first proposed by Dooley [Do], and used explicitly in
[DBE] and [CDB].

Let us first start by fixing some notations.  The contraction
adapted coordinates $(x, p, a, b, s)$ will describe $\e$ when
$x\in M_\k$ and $\ez$ when $x$ belongs to Minkowski
spacetime.  The presymplectic form on $\ez$ is given by
$\ome^0=-d\te^0$, where $\te^0$ is obtained from $\te$ in (3.9)
by contraction.  Actually, $\te^0=p\cdot dx + s\,b\cdot da$.  The
dot denotes here and throughout this section the flat metric scalar
product.

The kernel of $\ome^0$ is spanned by the vector fields $Y_H$
and $Y_{12}^0$ obtained by contraction of $Y_{50}$ and
$Y_{12}$ respectively.  More precisely,
$$Y_H\equiv\lk(\k Y_{50})=-{p\over
m}\cdotp{\partial\over\partial
x}\quad\hbox{and}\quad Y^0_{12}\equiv\lk
Y_{12}=b\cdotp {\partial\over\partial a}-a\cdotp
{\partial\over\partial b}.\eqno(6.1)$$
The vector fields $Y_H$ and $Y_{12}^0$ generate the right
action on $\ez$ of the subgroup $T\times SO(2)\subset\ppfl$,
where $T$ stands for the time translations subgroup.  The phase space
is then $\sz\equiv\ppfl/T\times SO(2)\cong\R^6\times S^2$.

When trying to prequantize as in 4.2, using $\ker\ome^0$, one is
faced with the following problem.  The space of $L^2$ functions
on $\ez$, such that,
$$(Y_H\psi)(x, p, a, b, s)=im\psi(x, p, a, b,
s)\enskip\hbox{and}\enskip(Y^0_{12}\psi)(x, p, a, b, s)=is\psi(x, p,
a, b, s),\eqno(6.2)$$
contains only the zero function.  This is a consequence of the
non-compact character of the subgroup generated by $Y_H$.  So
we will here proceed without requiring that (6.2) holds for $L^2$
functions on $\ez$.  A solution avoiding this problem will arise
subsequently.  As in section 4.1, the integrability condition
restricts $s$ to integer values, however $m\in\R_+$.

Let us now evaluate the \kz\ limit of the polarization vector
fields given in (4.5b),
$$\lk (\k Z_i)=i
Y_{P_i}\quad\hbox{where}\quad
Y_{P_1}=a\cdotp{\partial\over\partial x},\quad
Y_{P_2}=b\cdotp{\partial\over\partial
x}\quad\hbox{and}\quad Y_{P_3}={1\over
ms}s\cdotp{\partial\over\partial x}.\eqno(6.3)$$
The notation $Y_{P_i}$, originates from the fact that these
vector fields generate the right action of the space translations
subgroup of $\ppfl$ on $\ez$.  Note here that the complex character of the
vector fields $Z_i$ disappears when \kz.  For the spin part $\Xi$, the
contraction gives, $$\Xi^0\equiv\lk \Xi=-{2i\over
ms}\,s\cdotp{\partial\over\partial\zeta}+ims\,\bar\zeta\cdotp
{\partial\over\partial s}.\eqno(6.4)$$
The set of contracted vector fields $\{Y_H, Y^0_{12}, Y_{P_1},
Y_{P_2}, Y_{P_3}, \Xi^0\}$ spans a $\ppfl$-invariant
prepolarization on $\ez$.  In fact, algebraically, this is the
prepolarization obtained by Renouard [Re] in his quantization of
$\opoin$.  When projected on $\sz$, the latter produces a
$\ppfl$-invariant polarization which is neither K\"ahler nor real.
Its complex part $\Xi^0$ corresponds, as in the $\k\not=0$ case,
to the K\"ahler structure on the sphere $S^2$ in
$\sz\cong\R^6\times S^2$.  It is then natural to consider the UIR of $\ppfl$
that arises when using the previous prepolarization as the
\kz\ limit of $(\hmsk, \u)$.  In order to concretely identify that
representation, let us write down all the constraint equations that
the quantum states $\psi\in C^\infty(\ez)$ must satisfy.  The
approriate Hilbert space structure on these states will be considered later
on.  Taking into account the fact that $s^\mu$ can be expressed in
terms of $p$, $a$, and $b$ using $\epsilon_{\mu\nu\rho\sigma}p^\mu a^\nu b^\rho
s^\sigma=m^2s$, we first obtain for (6.2),
$$\eqalignno{(Y_H\varphi)(x, p, \zeta,
\bar\zeta)=im\varphi(x, p, \zeta,
\bar\zeta)\,&\Rightarrow\,-{p\over m}\cdotp{\partial\over\partial
x} \varphi(x, p, \zeta, \bar\zeta)=im\varphi(x, p, \zeta,
\bar\zeta)&(6.5a)\cr
(Y_{12}^0\varphi)(x, p, \zeta,
\bar\zeta)=is\varphi(x, p, \zeta,
\bar\zeta)\enskip
&\Rightarrow\,i(\zeta\cdotp{\partial\over\partial\zeta}-
\bar\zeta\cdotp{\partial\over\partial\bar\zeta}) \varphi(x, p, \zeta,
\bar\zeta)=is\varphi(x, p, \zeta, \bar\zeta)\qquad;&(6.5b)\cr}$$
and then, for the polarization conditions (as in 4.6), we obtain,
$$(Y_{P_1}\varphi)(x, p, \zeta, \bar\zeta)=0\,\Rightarrow\,
a\cdotp{\partial\over\partial x}\varphi(x, p, \zeta,
\bar\zeta)=0,\eqno(6.6a)$$
$$(Y_{P_2}\varphi)(x, p, \zeta, \bar\zeta)=0\,\Rightarrow\,
b\cdotp{\partial\over\partial x}\varphi(x, p, \zeta,
\bar\zeta)=0,\eqno(6.6b)$$
$$(Y_{P_3}\varphi)(x, p, \zeta, \bar\zeta)=0\,\Rightarrow\,
s\cdotp{\partial\over\partial x}\varphi(x, p, \zeta,
\bar\zeta)=0,\eqno(6.6c)$$
$$(\overline\Xi^0\varphi)(x, p, \zeta, \bar\zeta)=0\,\Rightarrow\,
s\cdotp{\partial\over\partial \bar\zeta}\varphi(x, p, \zeta,
\bar\zeta)=0.\eqno(6.6d)$$
Clearly conditions (6.5a) and (6.6a-c) fix the
$x$-dependence of $\varphi$ to be the phase factor $e^{i\,p\cdot
x}$.  Hence the general solution of the previous system is of the
following form,
$$\Phi(x, p, \zeta)=e^{ip\cdot x}\phi(p, \zeta)\quad \hbox{such
that}\quad \zeta\cdotp{\partial\over\partial\zeta}\phi(p, \zeta)=s\phi(p,
\zeta).\eqno(6.7)$$
The identification of the limiting $\ppfl$ representation is
simplified by the following observation.  The conditions (6.6a-c)
give rise, as conditions (6.5a-b), to infinitesimal unitary
characters of the one dimensional subgroups of space
translations.  They are clearly trivial.  Hence, the limiting
representation is nothing but the UIR of $\ppfl$ induced from the
unitary character $e^{i(m\tau+s\tau')}$ of the subgroup
$SO(2)\otimes_sT_{3,1}$.  Here $T_{3,1}$ stands for the
subgroup of spacetime translations and $\otimes_s$ denotes the
semi-direct product.  The representation space is then the space
of square integrable functions on $SO_0(3,1)/SO(2)$.

We actually have a realization of this representation.  In fact, the
Hilbert space $\hmsz$ is the space of $L^2$ functions on
$SO_0(3,1)$ satisfying (6.7).  The measure is the left Haar
measure on $SO_0(3,1)$.   Clearly, the phase factor $e^{ip\cdot
x}$ in (6.7) does not influence the square integrability, however
it is a crucial ingredient for the realization of the unitary action of
$\ppfl$ in $\hmsz$.  The generators of this action are explicitly obtained
through the contraction of the quantum observables $\hL$ given in (4.4).
Actually, $$\hat H\equiv\lk (\k\hat
L_{50})=i{\partial\over\partial x^0},\qquad \hat P_i\equiv\lk
(\k\hat L_{i5})=-i{\partial\over\partial x^i},\eqno(6.8)$$
and
$$\hat L^0_{\mu\nu} \equiv\lk\hat L_{\mu\nu}
=-i\Big\lbrack(x_\mu{\partial\over\partial
x^\nu}-x_\nu{\partial\over\partial x^\mu})+(x\rightarrow
p)+(x\rightarrow s)+(x\rightarrow a)+(x\rightarrow
b)\Big\rbrack,\eqno(6.9)$$
$\iu$ and $\mu, \nu \in \{0, 1, 2, 3\}$.  One then easily verifies
that $\hat H$, the $\hat P_i$'s and the $\hat L_{\mu\nu}$'s
realize the Poincar\'e Lie algebra.  The action of these operators
on the functions $\Phi$ in (6.7) is as follows,
$$\openup 1.5mm\eqalignno{(\hat H\Phi)(x,
p, \zeta)&=e^{ip\cdot x}(p^0\phi)(p, \zeta),&(6.10a)\cr
(\hat P_i\Phi)(x, p,
\zeta)&=e^{ip\cdot x}(p^i\phi)(p, \zeta),&(6.10b)\cr
(\hat
L^0_{\mu\nu}\Phi)(x, p, \zeta)&=-ie^{ip\cdot x}\Big\lbrack
(p_\mu{\partial\over\partial p^\nu}-p_\nu{\partial\over\partial
p^\mu})+(\zeta_\mu{\partial\over\partial
\zeta^\nu}-\zeta_\nu{\partial\over\partial \zeta^\mu})\Big\rbrack\phi(p,
\zeta);&(6.10c)\cr}$$
$\iu$ et $\mu, \nu \in \{0, 1, 2, 3\}$.

Let us summarize.  By contracting the UIR $(\hmsk, \u)$ of
$\so$ we obtain a UIR representation of $\ppfl$ by means of a
reinterpretation of the K\"ahler polarization conditions that
become real in the \kz\ limit.  In other words the holomorphic
induction becomes the usual induction.  In fact, the limiting
representation is induced from the unitary character
$e^{i(m\tau+s\tau')}$ of the subgroup $SO(2)\otimes_sT_{3,1}$,
which is trivial for the subgroup of space translations.  This
result is in perfect agreement with the one obtained by direct
geometric quantization of $\opoin$ [Re] (see also [Ra2]).

\bigskip

\noindent {\sect 7. Conclusions}
\medskip \noindent
In this concluding section let us first make some comments about
the limiting states obtained in (5.8) using the results of section 6.
Clearly, they are of the form (6.7).  In fact,
$$\Phi_{x', p', \zeta'}(x, p, \zeta)=e^{ip\cdot x}\left[
m^2p'^0\delta(\vec p-\vec p\,')\,e^{-ip'\cdot
x'}\left({\bar\zeta'\cdot\zeta\over
2}\right)^s\,\right].$$
One easily verifies that they are generalized eigenstates of $\hat H$ and
$\hat P_i$ and that they satisfy (6.10c).  They are generalized states
defined on $\hmsz$.  In particular, the state
$\Phi_0\equiv\Phi_{x', p_{(0)}, \zeta_{(0)}}$, for $p_{(0)}=(m,
0, 0, 0)$ and $\zeta_{(0)}=(0, 1, -i, 0)$, is a generalized
eigenstate of $\hat H$ and $\hat L_{12}^0$, with eigenvalue
$m$ and $s$, respectively.  All states of the form given above
belong to an orbit (of generalized
states on $\hmsz$) of the UIR obtained by contraction.  This orbit
is the zero curvature limit of $\of$.

The loss of the notion of optimal localization in the zero
curvature limit, reflects the non-existence of such a notion for a
$\ppfl$-invariant theory.  Moreover, it clearly arises as a
consequence of the break down of the K\"ahler charater of the
$\so$-invariant polarization when \kz.  In order to
recover this notion, one needs to introduce a fundamental
length, a positive constant curvature in our case.  This
observation confirms the regularizing role of the $\so$-invariant
theories as alternatives to the $\ppfl$-invariant ones.

\bigskip

\noindent{\sect Acknowledgements}
\medskip \noindent
The authors thank J.-P. Gazeau for countless stimulating discussions.  A. M.
E. thanks S.T. Ali and V. Hussin for their hospitality at Concordia University
and Universit\'e de Montr\'eal where this paper has been completed.

\bigskip
\noindent {\sect References}
\medskip
\parskip=2mm plus .5mm minus .5mm
\parindent=1.5cm

\refe{Ar}R. Arens, {\it Classical
Lorentz invariant particles\/}, J. Math. Phys. {\bf 12}, 2415-2422, 1971.

\refe{AIS}S. J. Avis, C. J. Isham and D. Storey, {\it Quantum field theory
in anti-de~Sitter space-time\/}, Phys. Rev. {\bf D18}, 3565-3576, 1978.

\refe{BEGG}R. Balbinot, M.A. El Gradechi, J.-P. Gazeau and B.
Giorgini, {\it Phase spaces for quantum elementary systems in
de~Sitter and Minkowski spacetimes\/}, J. Phys. A: Math. Gen.
{\bf 25}, 1185-1210,1992.

\refe{BFFS}B. Binegar, M. Flato, C. Fronsdal and S. Salam\'o, {\it
De~Sitter and conformal field theories\/}, Czech. J. Phys. {\bf B 32},
439-471, 1982.

\refe{BLL}H. Bacry and J.M. L\'evy-Leblond, {\it Possible kinematics\/}, J.
Math. Phys. {\bf 9}, 1605-1614, 1968.

\refe{CB}Y. Choquet-Bruhat, {\it Global solutions of Yang-Mills
equations on anti-de~Sitter spacetime\/}, Class. Quantum Grav. {\bf 6},
1781-1789, 1989.

\refe{CDB}C. Cishahayo and S. De Bi\`evre, {\it On the contraction of the
discrete series of $SU(1,1)$\/}, to appear in Ann. Inst. Fourier.

\refe{CPSW}M. Couture, J. Patera, R. T. Sharp and P. Winternitz, {\it
Graded contractions of $sl(3,\C)$\/}, J. Math. Phys. {\bf 32}, 2310-2318,
1991.

\refe{De}R. Delbourgo, {\it Minimal uncertainty states for the rotation
and allied groups\/}, J. Phys. A: Math. Gen. {\bf 10}, 1837-1846, 1977.

\refe{Di}W. G. Dixon, {\it Dynamics of extended bodies in general
relativity : I. Momentum and angular momentum\/}, Proc. Roy. Soc. London
{\bf A314},
499-527, 1970.

\refe{Do}A. H. Dooley, {\it Contractions of Lie groups and applications to
analysis\/}, in: Topics in modern harmonic analysis,
{\bf Vol. I}, 483-515, (Instituto Nazionale di Alta Matematica
Francesco Severi, Roma 1983).

\refe{DB1}S. De Bi\`evre, {\it Causality and localization in relativistic
quantum  mechanics\/}, in Proceedings of the conference Trobades
Scientifiques de la Mediterrania, 234-241, 1985.

\refe{DB2}S. De Bi\`evre, {\it Scattering in Relativistic Particle
Mechanics\/}, PhD thesis, University of Rochester, 1986.

\refe{DBE}S. De Bi\`evre and M. A. El Gradechi, {\it Quantum
mechanics and coherent states on the anti-de Sitter spacetime and their
Poincar\'e contraction\/}, to appear in  Ann. Inst. H. Poincar\'e: Phys.
Th\'eo., 1992.

\refe{DBEG}S. De Bi\`evre, M. A. El Gradechi and J.-P. Gazeau,
{\it Phase space description of a quantum elementary system on the
anti-de~Sitter spacetime and its contraction\/}; to appear in
Proceedings of the 18th ICGTMP, Moscow 1990.

\refe{DF}R. Delbourgo and J. R. Fox, {\it Maximum weight vectors possess
minimal uncertainty\/}, J. Phys. A: Math. Gen. {\bf 10}, L223-L235, 1977.

\refe{DS}W. De~Sitter, {\it On the relativity of inertia. Remarks
concerning Einstein's latest hypothesis\/}, Proc. K. Akad. Wet. Amsterdam
{\bf 19}, 1217-1225, 1917 and  {\it On the curvature of space\/},  Proc.
K. Akad. Wet. Amsterdam {\bf 20},
229-243, 1918.

\refe{E1}M. A. El Gradechi, {\it Th\'eories classique et quantique sur
l'espace-temps anti-de~Sitter et leurs limites \`a courbure nulle\/},
Th\`ese de Doctorat de l'Universit\'e Paris 7, Paris 1991 (unpublished).

\refe{E2}M. A. El Gradechi, {\it A geometric characterization of
masslessness for the anti-de~Sitter kinematics\/}, in preparation.

\refe{Ev}N. T. Evans, {\it Discrete series for the universal covering group
of the $3+2$ de~Sitter group\/}, J. Math. Phys. {\bf 8}, 170-184, 1967.

\refe{Fr1}C. Fronsdal, {\it Elementary particles in a curved space\/},
Rev. Mod. Phys. {\bf 37}, 221-224, 1965.

\refe{Fr2}C. Fronsdal, {\it Elementary particles in a curved space. II\/},
Phys. Rev. {\bf D10}, 589-598, 1974.

\refe{FH}C. Fronsdal and B. Haugen, {\it Elementary particles in a
curved space. III\/}, Phys. Rev. {\bf D12}, 3810-3818, 1975.

\refe{Gi}R. Gilmore, {\it Lie Groups, Lie Algebras and Some of Their
Applications\/}, (Wiley, New York, 1974).

\refe{GH}J.-P. Gazeau and M. Hans, {\it Integral-spin fields on
$(3+2)$-de~Sitter space\/}, J. Math. Phys. {\bf 29}, 2533-2552,1988.

\refe{H1}G. C. Hegerfeldt, {\it Remark on causality and particle
localization\/}, Phys. Rev. {\bf D10}, 3320-3321, 1975.

\refe{H2}G. C. Hegerfeldt, {\it Violation of causality in relativistic
quantum theory?\/} Phys. Rev. Lett. {\bf 54}, 2395-2398, 1985.

\refe{He}A. Heslot, {\it Observables et Structure Sympl\'ectique
en M\'ecanique Classique et en M\'ecanique Quantique\/}, Th\`ese de Doctorat
d'Etat, Universit\'e Paris VI, 1988.

\refe{Hu}N. E. Hurt, {\it Geometric Quantization in Action\/}, (D. Reidel
Publishing Company, 1983).

\refe{HE}S. W. Hawking and G. F. R. Ellis, {\it The Large Scale Structure of
Space-Time\/}, (Cambridge University Press, Cambridge 1973).

\refe{IW}E. In\"on\"u and E. P. Wigner, {\it On the contraction of groups
and their representations\/},  Proc. Nat. Acad. Sci. U. S. {\bf 39}, 510-524,
1953.

\refe{Ki1}A. Kirillov, {\it El\'ements de la Th\'eorie des
Repr\'esentations\/}, (Editions Mir, Moscou 1974).

\refe{Ki2}A. Kirillov, {\it The method of orbits in representation
theory\/}, in: Lie Groups
and Their Representations, I. M. Gelfand (Ed.), 219-230, (Akad\'emiai
Kiad\'o, Budapest 1975).

\refe{Ko}B. Kostant, {\it Quantization and unitary representations\/},
Lect. Not. Math. {\bf 170}, 87-207, (Springer-Verlag, New York 1970).

\refe{Ku}H. P. K\"unzle, {\it Canonical dynamics of spinning particles in
gravitational and electromagnetic fields\/}, J. Math. Phys. {\bf 13},
739-744, 1972.

\refe{LM}P. Libermann and C.M. Marle, {\it Symplectic Geometry and
Analytical Mechanics\/}, (D. Reidel Publishing Company, 1987).

\refe{LN}M. L\'evy-Nahas, {\it Deformation and contraction of Lie
algebras\/}, J. Math. Phys. {\bf 8}, 1211-1222, 1967.

\refe{M}L. Martinez-Alonso, {\it Group-theoretical foundations of
classical and quantum mechanics. II. Elementary systems\/}, J. Math. Phys.
{\bf 20}, 219-230,
1979.

\refe{Ma}G. W. Mackey, {\it On the analogy between semisimple Lie
groups and certain related semi-direct product groups\/}, in: Lie Groups
and Their Representations, I. M. Gelfand (Ed.), 339-364, (Akad\'emiai
Kiad\'o, Budapest 1975).

\refe{MN}J. Mickelsson and J. Niederle, {\it Contractions of
representations of de Sitter groups\/}, Comm. Math. Phys. {\bf 27},
167-180, 1972.

\refe{NW}T. D. Newton and E. P. Wigner, {\it Localized states for
elementary systems\/}, Rev. Mod. Phys. {\bf 21}, 400-406, 1949.

\refe{Pe}A. Perelomov, {\it Generalized Coherent States and their
Applications\/} (Springer Verlag, Berlin 1986).

\refe{PW}T. O. Philips and E. P. Wigner, {\it De~Sitter space and positive
energy\/}, in: Group Theory and Its Applications, E. M. Loebel (Ed.),
631-676, (Academic Press, New York 1968).

\refe{Ra1}J. H. Rawnsley, {\it De~Sitter symplectic spaces and their
quantizations\/}, Proc.Camb. Phil. Soc. {\bf 76}, 473-480, 1974.

\refe{Ra2}J. H. Rawnsley, {\it Representations of a semi-direct product
by quantization\/}, Math. Proc. Camb. Phil. Soc. {\bf 78}, 345-350, 1975.

\refe{Re}P. Renouard, {\it Vari\'et\'es Symplectiques et
Quantification\/}, Th\`ese Orsay, 1969.

\refe{Sa}J. Saletan, {\it Contraction of Lie groups\/}, J. Math. Phys.
{\bf 2}, 1-21, 1961.

\refe{Sc}W. Schmid, {\it $L^2$-cohomology and the discrete series\/},
Ann. Math. {\bf 103}, 375-394 (1976).

\refe{So}J. M. Souriau, {\it
Structure des Syst\`emes Dynamiques\/}, (Dunod 1970).

\refe{SM}E. C. G. Sudarshan and N. Mukunda, {\it Classical Dynamics: A
Modern Perspective\/}, (J. Wiley \& Sons, 1974).

\refe{SW}D. J. Simms and N. M. J. Woodhouse, {\it Lectures on Geometric
Quantization\/}, Lecture Notes in Physics {\bf 53}, (Springer-Verlag
1976).

\refe{W}A. S. Wightman, {\it On the localizability of quantum mechanical
systems\/}, Rev. Mod. Phys. {\bf 34}, 845-872, 1962.

\refe{Wi1}E. P. Wigner, {\it On unitary representations of the
inhomogeneous Lorentz group\/}, Ann. Math. {\bf 40}, 149-204, 1939.

\refe{Wi2}E. P. Wigner,  {\it  Some remarks on the infinite
de~Sitter space\/}, Proc. Nat. Acad. Sci. U. S. {\bf 36},
184-188, 1950.

\refe{Wo}N. M. J. Woodhouse, {\it Geometric Quantization\/},
(Clarendon Press, Oxford 1980).\end